\documentclass{pasj01}
\draft 
\Received{$\langle$reception date$\rangle$}
\Accepted{$\langle$acception date$\rangle$}
\Published{$\langle$publication date$\rangle$}
\usepackage{wrapfig}
\usepackage{color}
\usepackage{soul}
\usepackage{ulem}
\usepackage{lscape}
\usepackage[title]{appendix}
\usepackage{multirow}
\usepackage{setspace}
\twocolumn

\begin{document}

\title{Magnetic activity variability of nearby bright Sun-like stars by 4-year intensive H$\alpha$ line monitoring}
\author{
 Sanghee \textsc{Lee},\altaffilmark{1}
 Yuta \textsc{Notsu},\altaffilmark{2,3}
 and
 Bun'ei \textsc{Sato}\altaffilmark{1}
 }
 
   \altaffiltext{1}{Department of Earth and Planetary Sciences, Tokyo Institute of Technology,
   2-12-1 Ookayama, Meguro-ku, Tokyo 152-8551, Japan}
 \altaffiltext{2}{Laboratory for Atmospheric and Space Physics, University of Colorado Boulder, 3665 Discovery Drive, Boulder, CO 80303, USA}
 \altaffiltext{3}{National Solar Observatory, 3665 Discovery Drive, Boulder, CO 80303, USA}

\email{lee.s.ay@m.titech.ac.jp}

\KeyWords{stars:activity --- stars:chromospheres --- stars:solar-type --- planet–star interactions}

\maketitle

\begin{abstract}
We report intensive monitoring of the activity variability in the H$\alpha$ line for 10 Sun-like stars using the 1.88-m reflector at Okayama Branch Office, Subaru Telescope, during the last four years 2019-2022. Our aim was to investigate features of the stellar magnetic activity behaviors. We correlated the H$\alpha$ line variability of each star with the stellar activity levels derived from the Ca II H\&K line, suggesting its efficiency as a magnetic activity indicator.
In analyzing the H$\alpha$ line variation, we observed that some stars exhibited linear or quadratic trends during the observation period.
Among several G- and K-type stars expected to have co-existing activity cycles, we confirmed the 2.9-yr short cycle of $\epsilon$ Eri (K2V) from the H$\alpha$ observations. Additionally, we established upper limits on the H$\alpha$ variability of $\beta$ Com (G0V) and $\kappa$$^1$ Cet (G5V) concerning their expected shorter cycles. We also detected the possibility of short-term activity cycles in two F-type stars, $\beta$ Vir (F9V; $\sim$ 530 days) and $\alpha$ CMi (F5IV-V; $\sim$ 130 days). The cycle in $\alpha$ CMi was observed in only one season of our 4-yr observations, suggesting the temporal absence of the cycle period.
However, for stars with planets, we did not observe significant magnetic activity variability likely associated with the planetary orbital period. It is speculated that the impact of H$\alpha$ variability on radial velocity (RV) measurements may vary with spectral type.
\end{abstract}

\section{Introduction}
In Sun-like stars, the atmosphere exhibits stellar magnetic activity, generally induced by the magnetic fields and the active regions on the stellar chromosphere (\cite{linsky:2019}). 
The magnetic activity variability is a fundamentally important feature of the Sun-like stars because it is associated with the stellar magnetic activity and the dynamo contributed, which can also influence exoplanets.

The magnetic activity cycle is important to understand the behaviors of stellar magnetic activity, and continuous monitoring of the latest magnetic activity variability in different epochs using various chromospheric lines provides observational evidence to better understand the stellar magnetic activity. 
The long-term monitoring of the magnetic activity cycles in many Sun-like stars has primarily been carried out using the Ca II H\&K line (3933, 3968 \AA).
To date, most stars with well-determined cycles are concentrated in G- and K-type stars, displaying long and pronounced cycles ranging from 2.5 yr to more than 20 yr, as seen in the Mount Wilson HK program spanning several decades (\cite{baliunas:1995}).
For F-type stars, although there have not been many cases suggesting their activity cycle, recent research suggests that they have a relatively short activity cycle within 1-2 years (\cite{metcalfe:2010,mittag:2019}). Specially, the 123-day cycle of $\tau$ Bootis A (HD120136, F7V) has been reported through multi-wavelength observations (\cite{mittag:2017,lee:2023}). 

\citet{brandenburg:1998} suggested that there are two distinct relationships between the activity cycle and rotational rate, with an upper and a lower branch, 
as a particular activity behavior of the Sun-like stars, especially for G- and K-type stars. Considering the age and activity levels of stars on these two branches, young and active stars lie on the upper branch, and old and less active stars are positioned on the lower one within each spectral type category. 
In current activity cycles studies, it is proposed that some G- and K-type stars with co-existing cycles simultaneously occupy both branches, referred to as the ‘active branch’ and the ‘inactive branch, respectively (\cite{bohm:2007,brandenburg:2017}). 
A few conditions, such as stellar age or activity levels, have been suggested as potential factors for stars to exhibit co-existing cycles on the two branches, but without clear confirmation (\cite{baliunas:1995,brandenburg:2017}). It has been suggested that the stellar magnetic activity and its cycle may represent distinct features of stars depending on their spectral type. However, since most previous monitoring was conducted over long but relatively low time cadence, intensive monitoring over a few years would be necessary to elucidate activity variability features of F-, G-, and K-type stars, respectively. 

Meanwhile, the magnetic activity of the Sun-like stars is important to exoplanet searches because it can impact the detectability of exoplanets. 
The Radial velocity (RV) measurement is one of the best way to detect both long and short-term period exoplanets. However, it is also sensitive to stellar sources such as oscillations, granulation, and rotating active regions (\cite{lagrange:2010}).
Also, hot Jupiters can influence the activity variability of the parent stars through a magnetic reconnection between the stellar and the planetary magnetic field (known as star-planet magnetic interaction; SPMI) (\cite{cuntz:2000}; \cite{shkolnik:2008}).
Bright nearby Sun-like stars are the most promising targets for exoplanet detection, warranting follow-up observations using various methods.
A detailed investigation of long and short-term activity variability in nearby Sun-like stars while considering common characteristics such as spectral type or age is necessary.
Therefore, intensive monitoring with the proper cadence is also crucial in this context.

The magnetic activity variability of the Sun-like stars is possible to be traced by chromospheric lines other than the Ca II H\&K line, the most well-studied indicator. 
One of the activity indicators, the H$\alpha$ line (6563 \AA) would have lower amplitude variability than the Ca II H\&K line, so its long-term monitoring has not been sufficiently reported so far. 
Nevertheless, the H$\alpha$ line can be seen in the middle of the optical wavelength spectrum. Also, the H$\alpha$ line is more easily monitored due to its higher signal-to-noise ratio (S/N) than that of the Ca II H\&K line. 
As instruments such as CARMENES (wavelength coverage 5200-17100 \AA) for exoplanet searches, do not include the Ca II H\&K line spectrum, H$\alpha$ line observations become essential.
Moreover, it is suggested that the H$\alpha$ line is also useful for period determination as the spectroscopic periods measured from the Ca II H\&K and H$\alpha$ lines are in good agreement (\mbox{\cite{suarez:2015}}).
Therefore, the H$\alpha$ line can serve as a more accessible indicator for intensive monitoring, and offers valuable information for further observational studies of the stellar magnetic activity, such as the dependence on various stellar parameters. 

To investigate and provide insights into various aspects of magnetic activity behaviors, including activity cycles, amplitudes of activity variability, and exoplanet detection, we conducted intensive H$\alpha$ monitoring of 13 F-, G-, and K-type stars near the Sun over the past 4 years. These observations were carried out using the 1.88-m reflector at the Okayama Branch office of the Subaru Telescope. 
Among these stars, $\tau$ Boo, $\upsilon$ And, and $\tau$ Cet were previously reported in \citet{lee:2023}. 
In this paper, we present the H$\alpha$ line monitoring data for the remaining 10 stars.

The rest of this paper is organized as follows. 
The target stars are described in Section 2.
The observations and data reduction are described in Section 3. Analyses and results of the observed H$\alpha$ line are presented in Section 4, and we discuss the results in Section 5. Section 6 is devoted to a summary.

\section{Targets}
For our project, we carefully selected a total of 13 nearby bright Sun-like stars, which include $\tau$ Boo, $\upsilon$ And, $\beta$ Vir, $\alpha$ CMi, $\beta$ Com, $\kappa$$^1$ Cet, $\pi$$^1$ UMa, $\beta$ Aql, 61 Vir, $\tau$ Cet, $\epsilon$ Eri, $\eta$ Cep, and HD 102195.
The stellar parameters for all these target stars are summarized in Table \ref{tbl1}. 
In this paper, we focus on reporting our findings for 10 of these target stars, excluding $\tau$ Boo, $\upsilon$ And, and $\tau$ Cet. However, we include the stellar parameters of these three stars for the purpose of discussion.
The selected target stars exhibit a range of spectral types spanning F-, G-, and K-types, with ages ranging from 0.2 to 11.4 Gyr. Additionally, many of these stars have been previously studied in the context of exoplanet searches, with the existence of exoplanets already reported for $\tau$ Boo, $\upsilon$ And, 61 Vir, $\epsilon$ Eri, and HD 102195.
Among these 13 stars, five of them ($\tau$ Boo, $\epsilon$ Eri, $\beta$ Com, $\kappa$$^1$ Cet, and $\beta$ Aql) have been observed to display magnetic activity cycles in previous monitoring observations, as briefly summarized below. 

$\tau$ Boo, for instance, exhibits a four-month cycle as observed in the Ca II H\&K, X-ray, and H$\alpha$ observations (\cite{mittag:2017,lee:2023}).
$\epsilon$ Eri shows the co-existence of 2.9 and 12.7-yr cycles (\cite{metcalfe:2013}), with the 2.9-year cycle being observed during 2013-2022 (\cite{coffaro:2020,fuhrmeister:2023}, although its significance between 1985 and 1992 was reported as less clear (\cite{metcalfe:2013}).
$\beta$ Com demonstrates two cycles of 9.6 and 16.6 yr determined (\cite{baliunas:1995}), with the possibility of shorter cycles (2.8 and 14.5 yr) estimated by \citet{brandenburg:2017}. However, the shorter period might not have been accurately determined due to low monitoring cadence. 
$\kappa$$^1$ Cet, on the other hand, exhibits a single cycle of 5.6 yr (\cite{baliunas:1995}), although \citet{brandenburg:2017} suggests the presence of additional shorter cycles, with one estimated to be around 1 yr. Finally, $\beta$ Aql displays an activity cycle with a period of 4 yr (\cite{baliunas:1995}).

\onecolumn
\begin{landscape}
\begin{table}[h]
\caption{Basic parameters of target stars}\label{tbl1}
\begin{center}
\scalebox{1}[1]{ 
\begin{tabular}{ccccccccccc}\hline\hline
     & $\alpha$ CMi& $\tau$ Boo A & $\upsilon$ And A & $\beta$ Vir &   $\beta$ Com &  $\pi$$^{1}$ UMa \\
Parameter &HD61421 & HD120136 & HD9826  & HD102870  &  HD114710 & HD72905 &Reference\\
\hline			   			   
Spectral Type        & F5IV-V & F7V & F8V   & F9V &  G0V &G1.5V & \\
$V$ (mag)             & 0.37 & 4.50 & 4.10   & 3.6 &  4.26 &  5.63 & \\
$T_{\rm eff}$ (K)  &6530$\pm$50 &  6399$\pm$45& 6213$\pm$44  & 6132$\pm$26 & 5936$\pm$33 & 5884$\pm$6.8 & 1,2,3,4,4,5 \\ 
$\log g$ (cgs)   & 3.96 & 4.27$\pm$0.06 & 4.0$\pm$0.1 & 4.125$\pm$0.01 &  4.38 &  4.48& 1,2,3,6,7,8\\
$Age$ (Gyr) & 1.87$\pm$0.13 & 0.9$\pm$0.5& 3.12$\pm$0.12 & 2.9$\pm$0.3 & 1.7 &0.2 & 9,2,10,11,12,12\\
$R$ ($R_{\odot}$) & 2.048$\pm$0.025 & 1.42$\pm$0.08& 1.480$\pm$0.087  & 1.681$\pm$0.008 & 1.106$\pm$0.011 &0.95& 1,2,13,4,4,14\\
$M$ ($M_{\odot}$) &  1.499$\pm$0.031 & 1.39$\pm$0.25& 1.27$\pm$0.06 & 1.413$\pm$0.061 & 1.15 & 0.90& 9,2,10,6,15,14 \\
$v\sin i$ (km s$^{-1}$) & 3.16$\pm$0.50 & 14.27$\pm$0.06& 9.5$\pm$0.8 & 4.3 & 4.10$\pm$0.06& 14.27 & 1,2,10,16,17,18 \\
$\log R'_{HK}$& - & -4.759 & -5.04 & -4.94 &  -4.756 &-4.37& 19   \\
$P_{rot}$ (day) & 23 & 3.1$\pm$0.1  & 7.3$\pm$0.04 & 10 &   12.3$\pm$1.1& 5&20,21,22,20,23,24\\
Known short/long $P_{cyc}$ & - & $\sim$ 120 days/ - & - & - &  9.6/16.6 years& -& 21,23  \\
Proposed short/long $P_{cyc}$ (year)&- & - / $>$11 & - & - & 2.8/14.5 &- & 25,23  \\
\hline	
$P_{orb}$ (day) & - & 3.31 &4.62$\pm$0.23 & - & -& - & 26,27 \\
$M_{p}$ sin \textit{i} (M$_J$) & - &4.4 & 1.70 & - &  - & - & 28\\ 
$a$  (au) &  -&0.0462& 0.0594 &  - & - & - &28 \\   
          &  & ($\tau$ Boo b) & ($\upsilon$ And b)   &  &   & &\\

\hline
\end{tabular}
}
\end{center}

$^1$\citet{kervella:2004}, 
$^2$\citet{borsa:2015},
$^3$\citet{fuhrmann:1998}, 
$^4$\citet{boyajian:2012},
$^5$\citet{kovtyukh:2003}, 
$^6$\citet{north:2009},
$^7$\citet{mallik:1998},
$^8$\citet{soubiran:2008},
$^9$\citet{liebert:2013},
$^{10}$\citet{mcarthur:2010},
$^{11}$\citet{holmberg:2009},
$^{12}$\citet{mamajek:2008}, 
$^{13}$\citet{vanbelle:2009}, 
$^{14}$\citet{ribas:2005},
$^{15}$\citet{takeda:2007},
$^{16}$\citet{carrier:2005b}, 
$^{17}$\citet{gray:1997},
$^{18}$\citet{white:2007},
$^{19}$\citet{wright:2004a},
$^{20}$\citet{koncewicz:2007},
$^{21}$\citet{mengel:2016}, 
$^{22}$\citet{simpson:2010},
$^{23}$\citet{brandenburg:2017}, 
$^{24}$\citet{maldonado:2010},
$^{25}$\citet{baliunas:1996}
$^{26}$\citet{butler:2006},
$^{27}$\citet{piskorz:2017},
$^{28}$\citet{butler:1997}
\end{table}
\end{landscape}

\begin{landscape}
\begin{table}[h]
\addtocounter{table}{-1}
\caption{(Continued)}\label{}
\begin{center}
\scalebox{0.95}[0.95]{ 
\begin{tabular}{ccccccccccc}\hline\hline
     & $\kappa$$^1$ Cet& 61 Vir  &  $\beta$ Aql A & $\tau$ Cet & $\eta$ Cep& & $\epsilon$ Eri  & \\
Parameter &  HD20630  & HD115617  & HD188512 & HD10700 &  HD198149&HD 102195 & HD22049& Reference\\
\hline			   			   
Spectral Type    &  G5V  &G7V & G8IV   & G8V  &K0IV &K0V&K2V  \\
$V$ (mag)             & 4.84 & 4.74 &  3.71  & 3.50 & 3.41 & 8.07 &3.73   &   \\
$T_{\rm eff}$ (K)  &  5708  &  5338$\pm$13& 5071$\pm$37 & 5344$\pm$50 &4950&5283$\pm$29 &5084$\pm$5.9 & 8,29,30,31,32,5\\ 
$\log g$ (cgs)   &  4.51 &4.58&3.54$\pm$0.14& 4.44 &3.41& 4.53$\pm$0.03 & 4.30$\pm$0.08 & 8,33,34,35,31,32,36\\
$Age$ (Gyr) &  0.7 &  6.1-6.6 & 9.1-11.4 & 5.8 & 2.5$\pm$0.3 & 5.9$\pm$3.5 &0.4-0.8 & 12,37,32,38\\
$R$ ($R_{\odot}$) & 0.95$\pm$0.10 & 0.9867$\pm$0.0048 & 3.064$\pm$0.020 & 0.793$\pm$0.004 & 4.12$\pm$0.07 & 0.84$\pm$0.02&0.735$\pm$0.005& 39,29,40,31,32,41\\
$M$ ($M_{\odot}$) & 1.037$\pm$0.042 & 0.93 & 1.26$\pm$0.18  & 0.783$\pm$0.012 &1.6 & 0.88$\pm$0.03 &0.82$\pm$0.02 & 4,29,42,40,37,32,36 \\
$v\sin i$ (km s$^{-1}$) & 4.5&3.9$\pm$0.9& 22.28 & 0.90 & 6.79 & 2.6 &2.4$\pm$0.5 &43,44,44,45,46,47,48\\
$\log R'_{HK}$&  -4.454 & -4.96 & -5.21& -4.98 &-5.223 & -4.56 &-4.441 &  19,43,19,49,47,50\\
$P_{rot}$ (day) &  9.2$\pm$0.3 & 29 & 5.08697$\pm$0.00031 & 34 &  3.7 & 12.3 & 11.4 &43,51,52,23,53,54\\
Known short/long $P_{cyc}$ (year) & - /5.6 & - & - / $>$4&  - / $>$10 &  -& -& 2.9/12.9 & 25 \\
Proposed short/long $P_{cyc}$ (year) & 1/7.5 & - & - & - & - &  -& - /34  &23,55 \\
\hline	
$P_{orb}$ (day) &- & 4.21  &- &-&  - & 4.11 &6.85 years & 56,57,47\\
$M_{p}$ sin \textit{i} (M$_J$)  &- &0.019&  - &  -  & - &0.45 &1.55 & 56,57,47\\ 
$a$  (au) &- & 0.0502 & - &  - & - & 0.0491 &3.39 & \\   
        &   &(61 Vir b)  &  &  & - & (HD 102195 b)&($\epsilon$ Eri b)&  \\

\hline
\end{tabular}
}
\end{center}
$^{29}$\citet{vonbraun:2014}, 
$^{30}$\citet{santos:2004b}, 
$^{31}$\citet{piau:2011},
$^{32}$\citet{bonfanti:2016}, 
$^{33}$\citet{perrin:1988},
$^{34}$\citet{cenarro:2007}, 
$^{35}$\citet{cayrel:1992},
$^{36}$\citet{gonzalez:2010}, 
$^{37}$\citet{affer:2005},
$^{38}$\citet{janson:2015}, 
$^{39}$\citet{walker:2007}, 
$^{40}$\citet{teixeira:2009}, 
$^{41}$\citet{demory:2009},
$^{42}$\citet{bruntt:2010}, 
$^{43}$\citet{gaidos:2000},
$^{44}$\citet{ammler-voneiff:2012}, 
$^{45}$\citet{santos:2004a},
$^{46}$\citet{martinez-arnaiz:2010},
$^{47}$\citet{melo:2007},
$^{48}$\citet{frohlich:2007},
$^{49}$\citet{wittenmyer:2006}
$^{50}$\citet{henry:1996},
$^{51}$\citet{wyatt:2012},
$^{52}$\citet{butkovskaya:2018},
$^{53}$\citet{vidotto:2014},
$^{54}$\citet{roettenbacher:2022},
$^{55}$\citet{fuhrmeister:2023},
$^{56}$\citet{vogt:2010},
$^{57}$\citet{benedict:2006},

\end{table}
\end{landscape}

\twocolumn 
\section{Observation and data reduction}

\subsection{HIDES-F Observation}
We conducted spectroscopic observations of the target stars using the HIgh Dispersion Echelle Spectrograph (HIDES; \cite{izumiura:1999}) of the 1.88-m reflector at Okayama Branch Office, Subaru Telescope. The spectra were acquired by Fiber-Feed mode (HIDES-F; \cite{kambe:2013}), and we utilized high-efficiency mode with an image slicer. 
In 2018, HIDES was upgraded by removing the slit mode elements from the optical path, and rearranging the optical instruments on a new stabilized platform. The data used in this paper were obtained after the upgrade. 
The width of the sliced image was $\timeform{1.05''}$, corresponding to a spectral resolution of \textit{R} $\sim$ 55000 achieved by 3.8-pixel sampling. 
For our analysis, Table 2 lists the observation data of each target star reported in this work. We used one observed data per night. In the case of stars with several observations per night, we averaged the spectra for analyses.
The signal-to-noise ratio (S$/$N) for the stars is slightly different depending on weather conditions. 
For HIDES-F, the three CCDs cover a wavelength region of 3700-7500 \AA. However, there were severe aperture overlaps in a wavelength range of 3700-4000 \AA\ owing to the use of the image slicer and hence the scattered light for these apertures could not be subtracted. Consequently, we discard the overlapped apertures and did not analyze the Ca II H\&K lines for HIDES-F spectra.

\begin{table*}[h]
\caption{Observation data of target stars}\label{tbl2}
\begin{tabular}{ccccccccccc}\hline\hline
Target stars     &  $\beta$ Vir  &  $\alpha$ CMi & $\beta$ Com & $\kappa$$^1$ Cet & $\pi$$^{1}$ UMa  \\
\hline			   			   
observing nights &  110 & 160 &  136  & 87 &37   \\
observing period  &  2019.1-2022.5 & 2019.1-2022.4 &  2018.12-2022.6 & 2019.10-2022.1 & 2020.9-2022.5  \\
S/N (per pixel) &   100-150 &150-200 & 50-150 & 100-200 & 50-150  \\
exposure time (s) & 300 & 30 & 300 &  700 &700  \\
reference spectrum & 2019.5.2 & 2019.10.15 & 2019.1.28 & 2019.10.8 & 2021.12.14 
\end{tabular}
\begin{tabular}{ccccccccccc}\hline\hline
Target stars    &  $\beta$ Aql A & 61 Vir & $\epsilon$ Eri  & $\eta$ Cep& HD 102195 \\
\hline			   			   
observing nights   &  95  &  70  & 86 & 85 &71 \\
observing period  & 2019.10-2022.7 & 2019.2-2020.6 &2019.1-2022.6  & 2019.3-2022.7 &2020.4-2022.6  \\
S/N (per pixel)  &100-200 & 50-100 & 150-200  & 150-250 & 50-100 \\
exposure time (s) &  300 & 300 & 350 & 300 & 1800 \\
reference spectrum  & 2019.10.15 &  2019.3.4&2019.1.28 &2019.5.16&2020.4.29 \\
\hline	
\end{tabular}
\end{table*}

\subsection{Echelle data reduction and telluric correction}
The echelle data reduction was performed in a standard manner using the IRAF\footnote{IRAF is distributed by the National Optical Astronomy Observatories, which is operated by the Association of Universities for Research in Astronomy, Inc. under a cooperative agreement with the National Science Foundation, USA} packages to extract one-dimensional raw spectra (bias subtraction, flat fielding, scattered-light subtraction and spectrum extraction). 
The wavelength calibration was determined from Thorium-Argon arcs taken before and after each spectrum. 
For each night spectrum, we performed the correction for telluric absorption lines using $\alpha$ Leo (B8IV, \textit{v} sin \textit{i} = 317 km s$^{-1}$), which was taken on the same night as a standard star. 
We also corrected the dependence on airmass, which varies with the elevation of the object, by using the standard star spectrum (\cite{kawauchi:2018}).
The continuum-normalization for the telluric-line-corrected spectra was carried out with a reference spectrum in Table \ref{tbl2} (see \cite{lee:2023}).

\begin{figure}[h]
 \begin{center}
  \includegraphics[width=80mm]{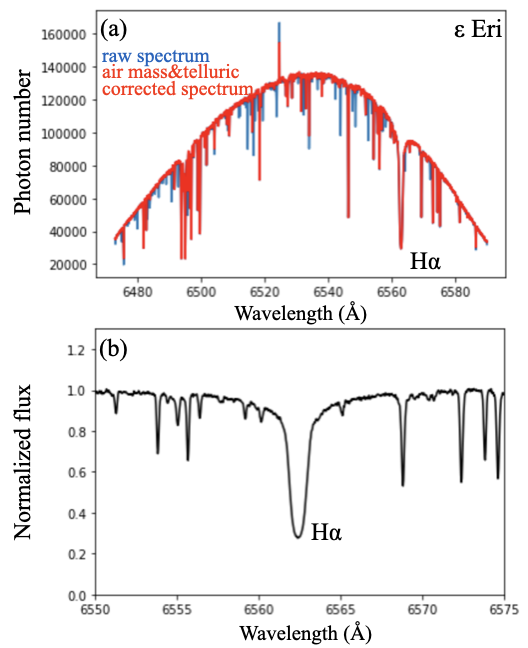}
 \end{center}
 \caption{(a) Single raw spectrum (blue) and telluric line corrected spectrum (red) of H$\alpha$ line for $\epsilon$ Eri. (b) Single continuum-normalized spectrum of H$\alpha$ line for $\epsilon$ Eri.}\label{fig1}
\end{figure}

We show the result of the data reduction and telluric correction with $\epsilon$ Eri. Figure \ref{fig1} (a) shows the single raw spectrum and the telluric-line-corrected spectrum for $\epsilon$ Eri obtained after the data reduction. Figure \ref{fig1} (b) shows the normalized H$\alpha$ line profile and its core for $\epsilon$ Eri.
The entire procedure was performed in the same manner for all target stars.

\section{Analyses}
\subsection{H$\alpha$ line intensity}
\begin{figure*}[h]
 \begin{center}
  \includegraphics[width=111mm]{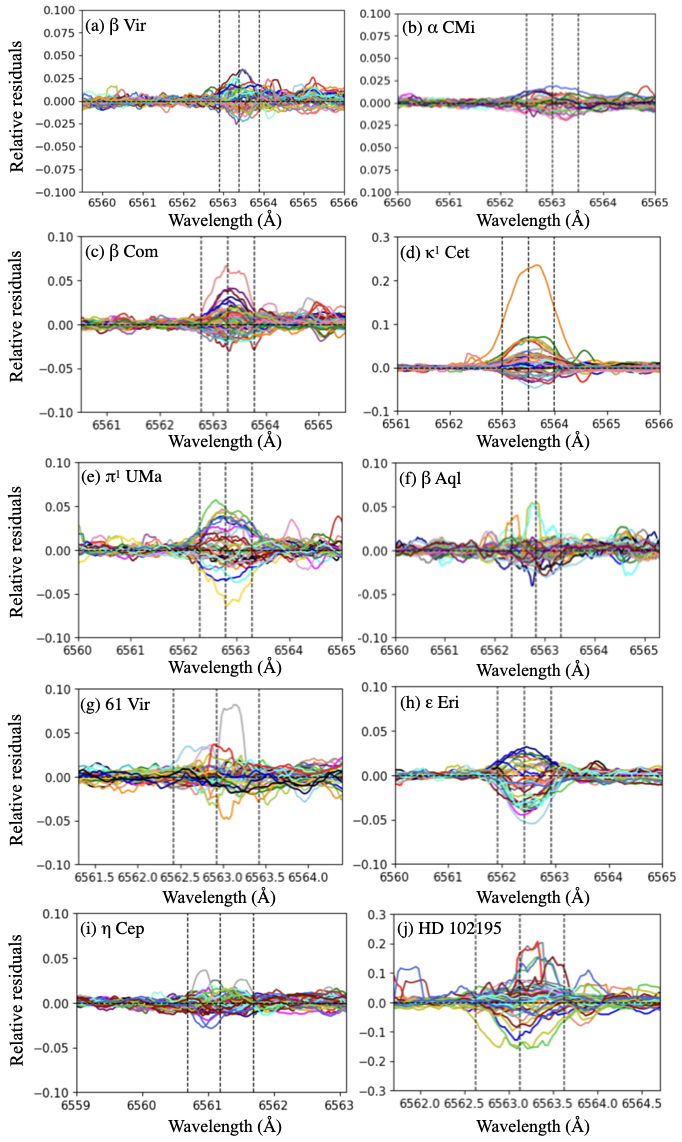}
 \end{center}
 \caption{Residuals relative to the average spectrum in H$\alpha$ lines of the target stars (a part of all spectra). The residuals have been smoothed to avoid cluttering. The dashed lines indicate a range of $\sim$ 1\AA\ around the line core.}\label{fig2}
\end{figure*}

To quantify the variability strength in the H$\alpha$ line core for each star, we averaged all the continuum-normalized spectra and then calculated the relative residuals for each normalized spectrum in relation to the averaged spectrum.
Figure \ref{fig2} shows the relative residuals of the target stars (part of all spectra), which have been smoothed to show the variability in the H$\alpha$ line core more clearly. Each star exhibits nightly variability in the H$\alpha$ line core throughout the observing seasons.

\begin{figure*}[h]
 \begin{center}
  \includegraphics[width=111mm]{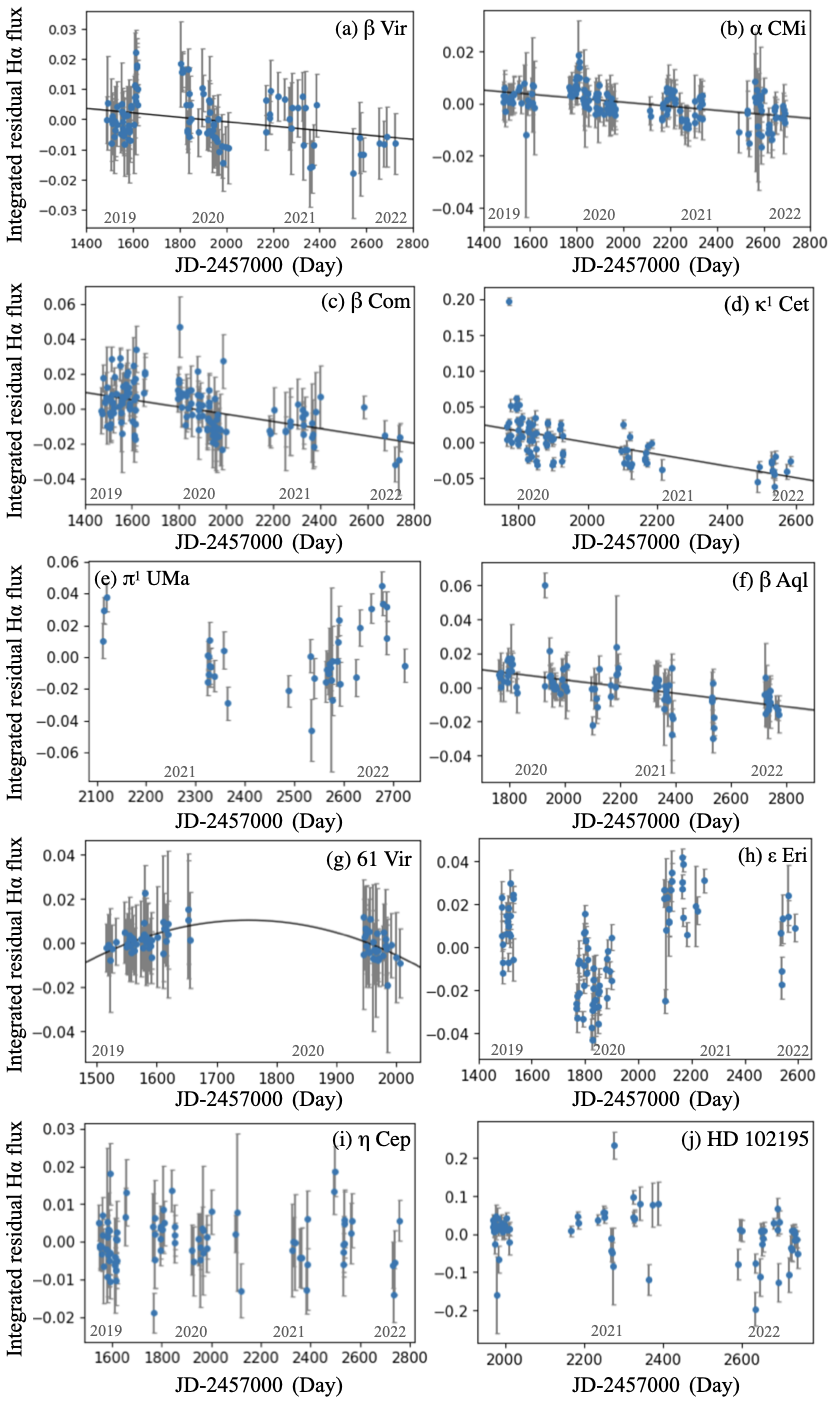}
 \end{center}
 \caption{Integrated residual H$\alpha$ flux of the target stars. The error bars in the integrated residual flux are calculated by the intranight photon noise. For several stars, the fit of the trends found by F-test is also depicted. 
 }\label{fig3}
\end{figure*}

To evaluate the level of the H$\alpha$ variability, we integrated the relative residuals over a $\sim$ 1 \AA\ interval centered around the line core. This is the same method as described in \citet{borsa:2015}, which observed one of our target star $\tau$ Boo. 
In Figure \ref{fig3}, we plot the integrated residual H$\alpha$ flux along the time series of the target stars to present the intensity of the H$\alpha$ line core.
Error bars represent the random noise associated with photon fluxes captured by the spectrograph. 

Each of the target stars has exhibited distinct variations in the H$\alpha$ intensity.
$\epsilon$ Eri shows an extremely clear variation throughout the time series (Fig. \ref{fig3} (h)).
Also, several stars such as $\beta$ Vir, $\alpha$ CMi, and $\pi$$^{1}$ UMa, imply periodic variations (Fig. \ref{fig3} (a),(b),(e)).
The remaining stars show a gradual decrease in the integrated residual flux, implying long-term trends.
$\kappa$$^1$ Cet shows some outlier data points (Fig. \ref{fig3} (d)).
Considering that this star is known to exhibit flares (\cite{hamaguchi:2023}), and since other stars did not show similar noise pattern during nearly the same period, the outliers may be derived from sudden increase in intrinsic activity.
We obtained the variability of the integrated residual H$\alpha$ flux (hereafter called `H$\alpha$ variability') by calculating the rms of the amplitude for each star, which ranges from 0.9\% to 9.4\%.

\subsection{Long and short-term variations in the H$\alpha$ line}
We determined the period of the integrated residual H$\alpha$ flux and peaks significance, using the generalized Lomb-Scargle (GLS) method and the false alarm probability (FAP) by \mbox{\citet{zechmeister:2009}}.
Given that the minimum data point interval is approximately one day and our focus on periodic variations longer than the rotational periods of the target stars ($>$ 3-5 days), each periodogram was calculated up to its Nyquist frequency of 0.5 day$^{-1}$. 
We also estimated the error range of the determined period by using normal distribution random numbers within the error bars for each data point of the integrated residual H$\alpha$ flux.
We generated 10$^{4}$ data sets, choosing random values corresponding to the error bars at each point in one data set. Then, we calculated the GLS periodogram to estimate the error range. 

For stars exhibiting a visible long-term trend in the integrated residual H$\alpha$ flux, we assessed the presence of actual long-term trends, including linear or quadratic trends. To address the significance of the visible trend, we used the F-test in a similar way to that described by \citet{cumming:1999} (see also \citet{lee:2023} for details). The results of the F-test are described with the FAP, that here indicates pure noise fluctuations would produce a trend by chance. The found trends for several stars are depicted in Figure \ref{fig3}, alongside the integrated residual H$\alpha$ flux .
Long-term trends might influence the power spectrum of the periodogram, so we checked detrended data in the periodogram after removing the linear or quadratic long-term variations.
For stars without a long-term trend found by the F-test, we removed possible periodicity. 

We also investigated significant peaks in the short-term period range close to stellar rotational or planetary orbital period.
We confirmed that the HIDES H$\alpha$ observations show stellar intrinsic activity, not instruments, seasons, or any non-stellar-induced factors, by comparing them to  observations of an inactive star, $\tau$ Cet (\cite{lee:2023}).

Additionally, we adopted a so-called bootstrap randomization method to estimate detection limits for possible periodicity. 
We generated noise data sets by redistributing the actual fit residuals for each star. We then injected a fake data set of the integrated residual H$\alpha$ flux with a fixed period and certain amplitude as a signal against the noise. We calculated the GLS periodogram, examined the power of the peak at the period, and regarded it as detection if the power exceeded the level of FAP=0.1\%. We repeated the procedure for 10$^{5}$ fake data sets, and counted the number of the detection, which was regarded as probability of the detection for the signal with the injected period and the amplitude. We performed this simulation for various amplitudes and set the amplitude with the detection probability of 99\% to be the detection limit at the period.

\section{Results and Discussion}
In this section and the appendix, we present periodograms of the target stars (Fig. \ref{fig5},\ref{fig7},\ref{figA1_1}). The results of the detrended data are also shown in the periodogram. In the window function, the highest peaks correspond to an observation sampling of one year (f = 0.00265 day$^{-1}$), and its aliases and other noise in the power spectrum appear as lower peaks. Additionally, Table \ref{tbl3} provides details on the results of the F-test, the H$\alpha$ variability, existence of any periodicity associated with activity cycles, stellar rotational period, or planetary orbital period, and the detection limits for the period. 

\begin{table*}[h]
\caption{Results of periodogram analyses of target stars}\label{tbl3}
\begin{tabular}{cccccccccccc}\hline\hline
Target stars     & $\beta$ Vir  &  $\alpha$ CMi & $\beta$ Com & $\kappa$$^1$ Cet & $\pi$$^{1}$ UMa  \\
\hline			   			   
long-term trend by F-test  &  linear & linear &  linear   & linear & -    \\
FAP of the trend  &  $\approx$ 10$^{-3}$ & $<$10$^{-5}$ &  $<$10$^{-5}$ & $<$10$^{-5}$ & - \\
H$\alpha$ variability (\%) & 1.3 & 0.9 & 1.7 & 2.5 & 2.4  \\
Periodicity with FAP $<$ 0.1\% (days)  & 530 & 130& - & - & 560 \\
Peaks in short-term range close to & \multirow{2}{*}{-} & \multirow{2}{*}{-}& \multirow{2}{*}{12.3} & \multirow{2}{*}{9.3} & \multirow{2}{*}{4.8}  \\
P$_{rot}$ or P$_{orb}$ (days)  & & & & &  \\
\end{tabular}

\begin{tabular}{cccccccccccc}\hline\hline
Target stars    &  $\beta$ Aql A & 61 Vir & $\epsilon$ Eri  & $\eta$ Cep& HD 102195 \\
\hline			   			   
Long-term trend by F-test   &  linear  & quadratic  & - & - &- \\
FAP of the trend  & $<$10$^{-5}$ & 8$\times$10$^{-5}$ &- & - &-  \\
H$\alpha$ variability (\%) & 1.5 & 1.9&  2.1 & 1.13 & 9.4\\
Periodicity with FAP $<$ 0.1\% (days) & - &- & 530, 1060 & - & - \\
Peaks in short-term range close to & \multirow{2}{*}{-} & \multirow{2}{*}{37} & \multirow{2}{*}{11.4}& \multirow{2}{*}{-} & \multirow{2}{*}{-} \\
P$_{rot}$ or P$_{orb}$ (days)  & & & & &  \\

\hline	
\end{tabular}
\end{table*}

\subsection{Dependence of the H$\alpha$ variability on stellar parameters}
\begin{figure*}[h]
 \begin{center}
  \includegraphics[width=140mm]{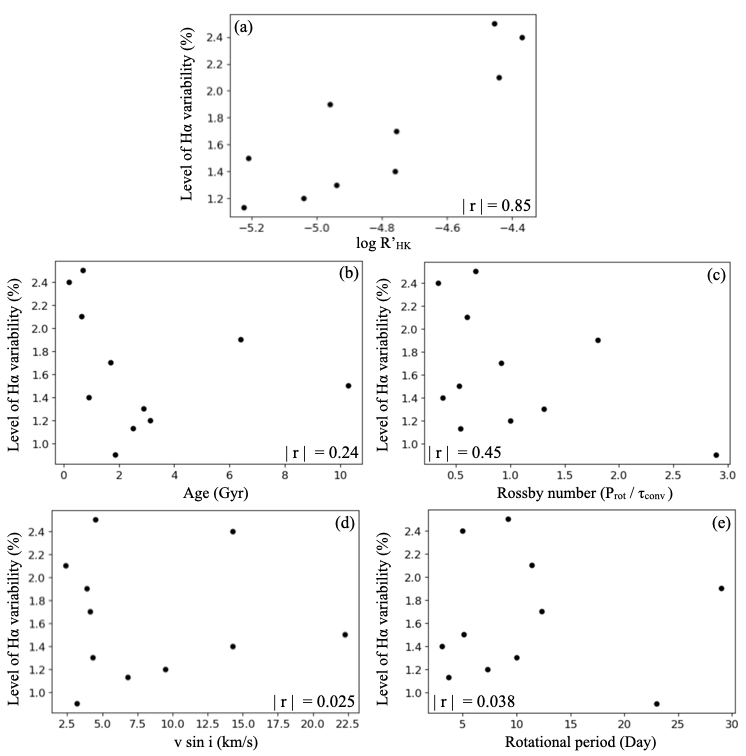}
 \end{center}
 \caption{Correlation between (a) log R'$_{HK}$, (b) stellar age, (c) Rossby number, (d) rotational velocity (\textit{v} sin \textit{i}), (e) rotational period, and the level of the H$\alpha$ variability. The correlation coefficient (r) is presented in each figure.}\label{fig4}
\end{figure*}

With the obtained H$\alpha$ variability, we investigated the possible relationship between the H$\alpha$ variability of individual stars and their stellar parameters. We plot the H$\alpha$ variability as a function of activity level in Figure \ref{fig4} (a). The activity level is quantified as the value of log R'$_{HK}$, which is derived from R'$_{HK}$ defined as the ratio of chromospheric emission in the cores of the Ca II H\&K line to the total bolometric emission of the star (\cite{noyes:1984}). 

In Figure \ref{fig4} (a), we found a significant correlation (correlation coefficient (r) = 0.85) between the H$\alpha$ variability and activity level. More active stars exhibit greater variability in the H$\alpha$ line. 
It is widely accepted that more active stars show greater activity variability in the Ca II H\&K line. Our result suggests a similar dependence in the H$\alpha$ line.
The strong correlation between the H$\alpha$ variability and log R'$_{HK}$, which indicates the Ca II H\&K core emission flux, suggests that the H$\alpha$ line observed with HIDES-F can effectively trace activity variability in Sun-like stars, similar to the Ca II H\&K line.

We also plotted the H$\alpha$ variability as a function of several other parameters: stellar age, Rossby number, and rotational velocity (\textit{v} sin \textit{i}) in Figure \ref{fig4}.
In Figure \ref{fig4} (b), a very weak correlation was found between stellar age and the H$\alpha$ variability (r = -0.24). Also, a weak correlation was implied between the Rossby number and H$\alpha$ variability (r = -0.45) (Fig. \ref{fig4} (c)). 
We could not find any significant correlation between rotational velocity and the H$\alpha$ variability, or between rotational period and the H$\alpha$ variability (Fig. \ref{fig4} (d)\&(e)).
It suggests the possibility of a weak relationship between stellar age or Rossby number and the H$\alpha$ variability.
The stars with similar activity effects by similar Rossby numbers show varying degrees of activity variability.

For the Sun-like stars, previous studies have explored the relationship between the Ca II H\&K emission and stellar age, mass, or color index
(\cite{wright:2004b,lorenzo-oliveira:2018,linsky:2019}). As we did not find a similar dependence with the log R'$_{HK}$ for several parameters of the target stars (Fig \mbox{\ref{figB1}}), it is evident that more cases would be necessary to establish a clear relationship between activity variability and stellar parameters.

\subsection{Magnetic activity cycle and behavior in the H$\alpha$ line for individual stars}
 \subsubsection{$\epsilon$ Eri}
\begin{figure*}[t]
 \begin{center}
  \includegraphics[width=150mm]{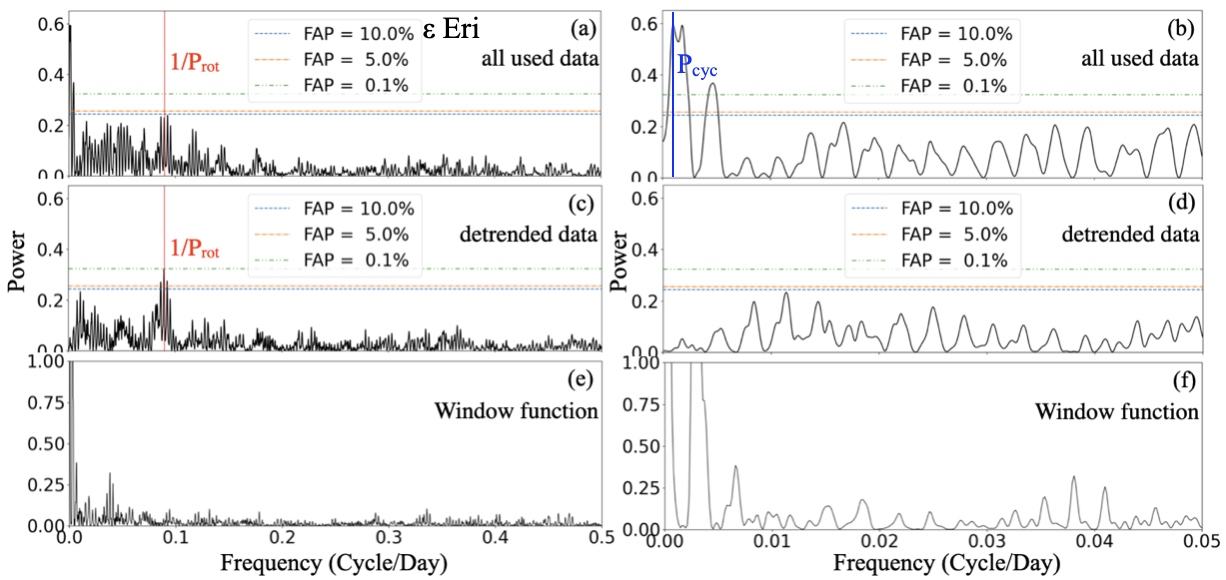}
 \end{center}
 \caption{(a) Periodogram of the integrated residual H$\alpha$ flux time series of $\epsilon$ Eri. (b) Enlarged portion of (a). (c) Periodogram of the detrended H$\alpha$ flux data of $\epsilon$ Eri. (d) Enlarged portion of (c). (e) Window function of the H$\alpha$ flux time series of $\epsilon$ Eri. (f) Enlarged portion of (e). The red line represents the stellar rotational period. The blue line represents the short activity cycle from the previous report (\cite{metcalfe:2013}).}\label{fig5}
\end{figure*}

In Figure \ref{fig5} (a)\&(b), we identified the most significant peak at 1060 days (f = 0.0009 day$^{-1}$) with FAP $<$ 0.1\%. However, the 530 days (f = 0.00018 day$^{-1}$) was also significant, exhibiting a power level comparable to the 1060-day periodicity. In the short-term range, a peak at 11.4 days (f = 0.0877 day$^{-1}$), which is close to the stellar rotational period, was implied. After removing the long-term periodicity, the power of the 11.4-day peak increased.

\begin{figure*}[t]
 \begin{center}
  \includegraphics[width=150mm]{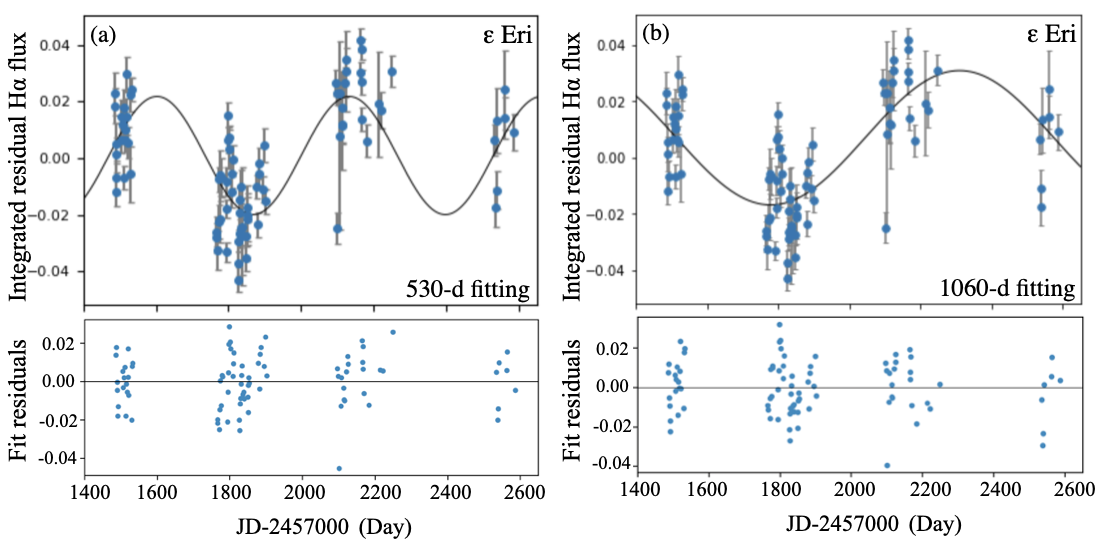}
 \end{center}
 \caption{(a) Integrated residual H$\alpha$ flux time series (upper panel) and the fit residuals (lower panel) with the sinusoidal fit of 530 days for $\epsilon$ Eri. (b) Integrated residual H$\alpha$ flux time series (upper panel) and the fit residuals (lower panel) with the sinusoidal fit of 1060 days for $\epsilon$ Eri. The error bars in the integrated residual flux are calculated by the intranight photon noise.}\label{fig6}
\end{figure*}

We present sinusoidal curves fitted to the integrated residual H$\alpha$ flux for the 530-day and 1060-day (Fig. \ref{fig6} (a)\&(b)). While the 530 days is the most likely alias of the 1060 days, both periodicities provide a good fit based on the sinusoidal curve results. 
Thus, we used the bootstrap randomization to validate the two significant periodicities. 
The periodogram showed peaks at $\sim$ 1060 days and 530 days, each with a FAP well below 0.1\%, but only when we injected a signal with a period of $\sim$ 1060 days and an amplitude similar to or greater than the observed one. The 1060 days did not appear as a significant peak with FAP $<$ 0.1\% when we injected a signal with a period of $\sim$ 530 days, which differs from our H$\alpha$ result. Thus, we conclude that the 1060-day period represents the true activity cycle of $\epsilon$ Eri.
The error range of the detected period is 1060 $\pm$ 38 days.

The presence of a possible $\sim$ 2.9 years ($\sim$ 1060 days) magnetic activity cycle in our periodogram of $\epsilon$ Eri is precisely consistent with the previous Ca II H\&K observations and reveals a shorter period cycle in the star. 
Though it was suggested that the shorter cycle might not persist temporarily by the previous analysis of the archival data between 1985 and 1992 (\cite{metcalfe:2013}), our H$\alpha$ result also shows the 2.9-yr cycle, which has remained consistent in same phase.

Given the previous Ca II H\&K observation of the 2.9-yr cycle for $\epsilon$ Eri (\cite{coffaro:2020}), one might expect the H$\alpha$ variability, inferred from the relative amplitude of the Ca II H\&K line (S$_{MWO}$), to be at least 6.0\% or higher. However, we detected the 1060-day cycle with the 2.1\% H$\alpha$ variability. This suggests that the activity cycle in $\epsilon$ Eri may be detected even with smaller amplitudes than the Ca II H\&K variability.

Meanwhile, \citet{fuhrmeister:2023} extended the time series of the S$_{MWO}$ data, covering more than 50 years, and proposed a 34-yr cycle in addition to the previously known 2.9 and 12.7-yr cycles. For $\epsilon$ Eri, the detection of the rotational period depends on longer trends, differential rotation, or individual active regions on timescales shorter than the rotational period due to its high activity.
It has been suggested that the time intervals during which a significant peak at the rotational period of this star was detected roughly correspond to the late decay phase and minimum of the 34-yr activity cycle (\cite{fuhrmeister:2023}).
Additionally, the Ca II IRT line, which exhibits much lower amplitude than the Ca II H\&K also showed a significant peak at the rotational period in the same time intervals (\cite{fuhrmeister:2023}).
In our periodogram, the rotational period was significantly detected, and this 4-yr H$\alpha$ observation clearly reveals the shorter cycle of $\epsilon$ Eri without implying the existence of a longer cycle. The absence of the longer cycle is explained, since our data covers the minimum of the longer cycle.

\subsubsection{$\alpha$ CMi}
\begin{figure*}[h]
 \begin{center}
  \includegraphics[width=150mm]{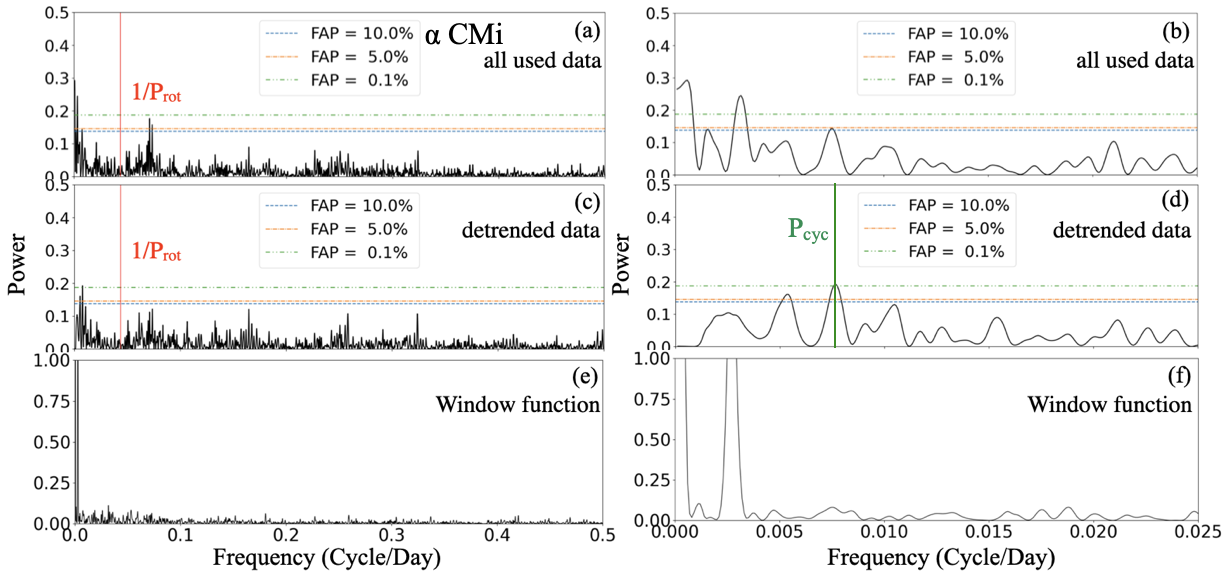}
 \end{center}
 \caption{(a) Periodogram of the integrated residual H$\alpha$ flux time series of $\alpha$ CMi. (b) Enlarged portion of (a). (c) Periodogram of the detrended H$\alpha$ flux data of $\alpha$ CMi. (d) Enlarged portion of (c). (e) Window function of the H$\alpha$ flux time series of $\alpha$ CMi. (f) Enlarged portion of (e). The red line represents the stellar rotational period. The green line represents the short activity cycle found in the integrated residual H$\alpha$ flux.}\label{fig7}
\end{figure*}

For $\alpha$ CMi, the periodogram shows significant peaks with FAP $<$ 0.1\% at 318 days (f = 0.00314 day$^{-1}$) (Fig. \ref{fig7} (a)\&(b)). A long-term variation (f $<$ 0.0006 day$^{-1}$), possibly associated with the linear trend, was found in the periodogram as well. After removing the linear trend, a significant peak at 130 days (f = 0.00769 day$^{-1}$) with FAP $\approx$ 0.1\% was found (Fig. \ref{fig7} (c)\&(d)). The error range of the detected period is 129.6 $\pm$ 6.2 days.

\begin{figure*}[h]
 \begin{center}
  \includegraphics[width=160mm]{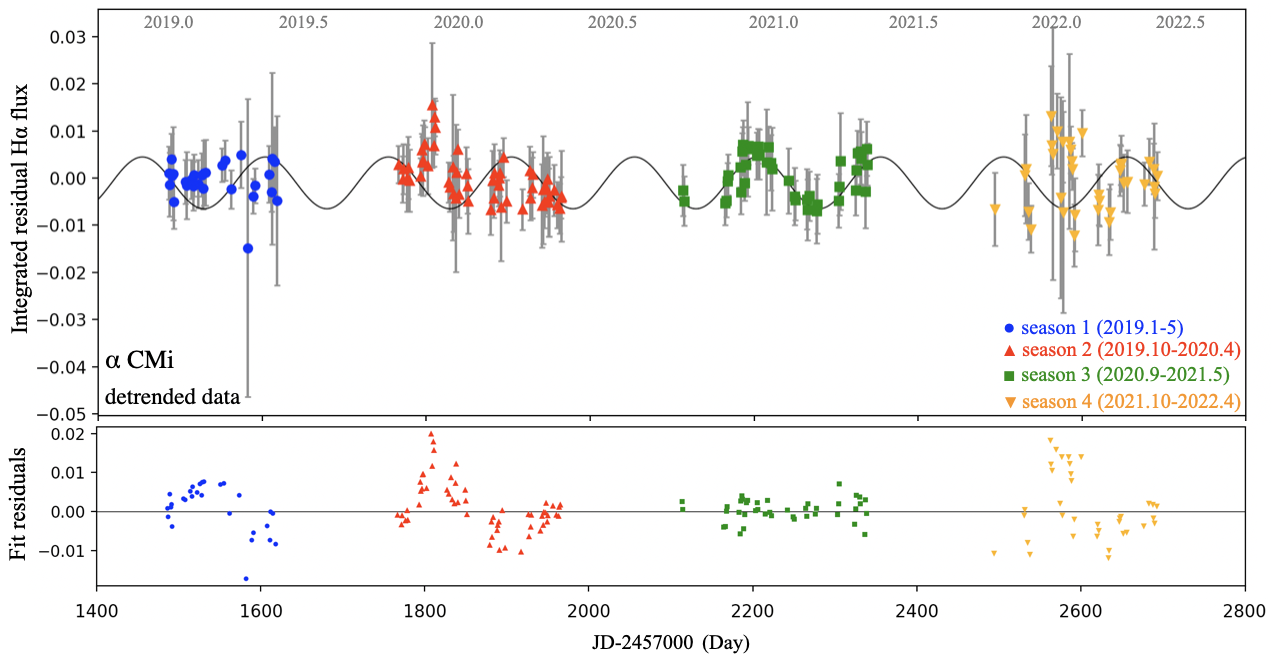}
 \end{center}
 \caption{Detrended data of the integrated residual H$\alpha$ flux time series (upper panel) and the fit residuals (lower panel) of $\alpha$ CMi. The solid line depicts the sinusoidal fit with a period of 130 days. The different colors and symbols represent the each observing season during the four years (2019-2022). The error bars in the integrated residual flux are calculated by the intranight photon noise.}\label{fig8}
\end{figure*}

\begin{figure*}[h]
 \begin{center}
  \includegraphics[width=160mm]{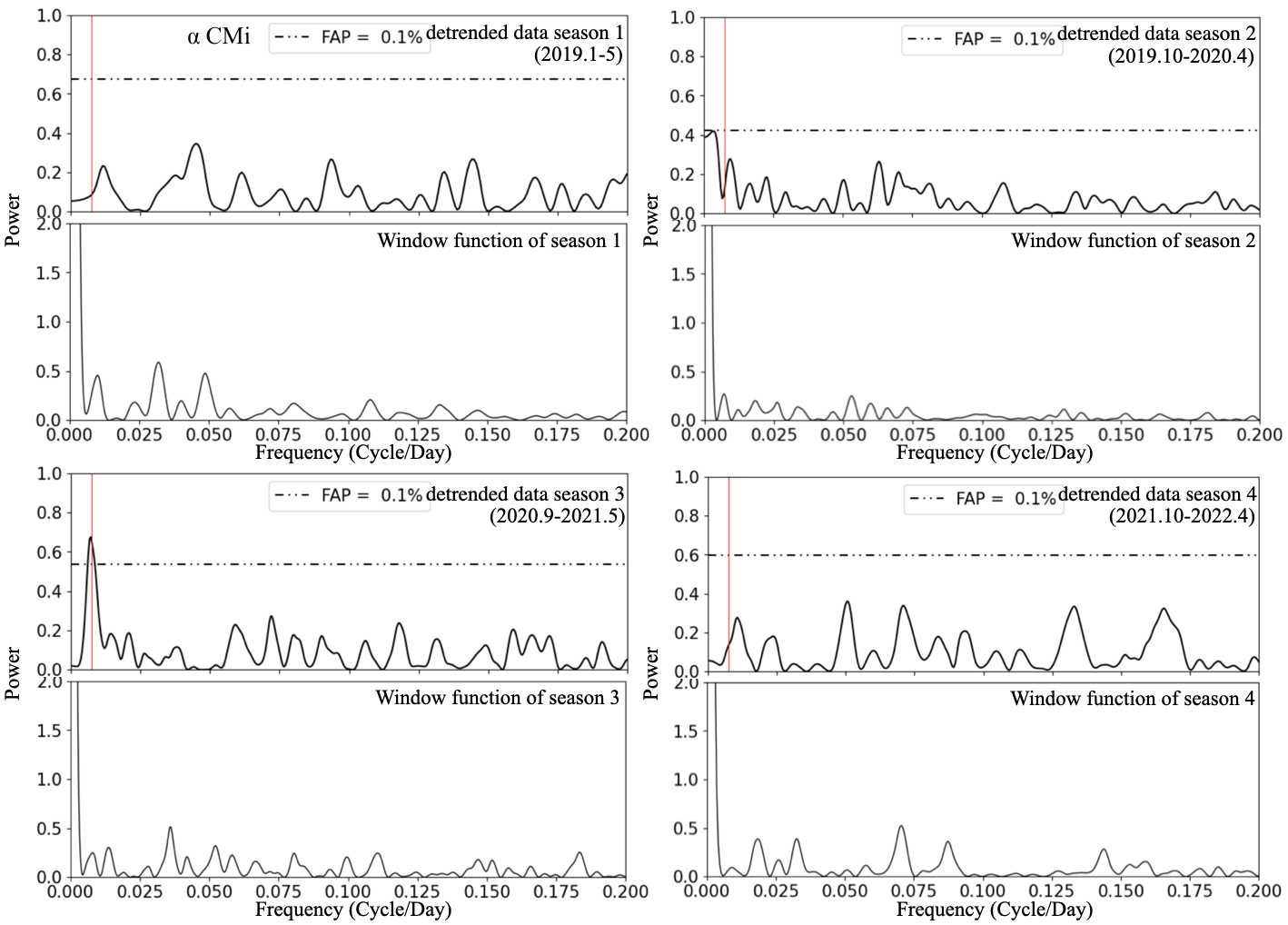}
 \end{center}
 \caption{Periodogram of the seasonal H$\alpha$ flux time series of $\alpha$ CMi (upper panel) and window function of the seasonal H$\alpha$ flux time series of $\alpha$ CMi (lower panel) for each season. The red lines represent the period of 130 days from the result of the detrended data of the entire time series.}\label{fig9}
\end{figure*}

The peak at 130 days may be influenced by the visible variation in season 3 (2020.9-2021.5) rather than the long-term trend, as shown in Figure \ref{fig8}. 
To explore this further, we analyzed individual observing seasons separately. The results are shown in Figure \ref{fig9}. Season 3 shows a distinct period of $\sim$ 130 days with FAP = $4 \times 10^{-6}$. Such a peak at 130 days was not found in other stars observed with HIDES-F in almost same seasons. 
However, season 1,2, and 4 did not show a specific period of $\sim$ 130 days. Season 2 exhibits a relatively significant peak at 280 days (f = 0.00357 day$^{-1}$) with FAP = 0.1\%.

The observing span of season 1 does not cover the 130 days, however, those of season 2 and 4 seem to be sufficient to detect the period. 
To determine whether the absence of the 130-day period in other individual seasons is due to a lack of data or the non-existence of such a period, the simulation analysis was performed on 10$^{5}$ fake data sets with the same period and amplitude as season 3 along the times series of season 1, 2, and 4, respectively (see section 4.2 for the bootstrap method details). We estimated the FAP of the 130-day period by obtaining the frequency at which the period exceeded FAP = 0.1\% in the periodogram of the H$\alpha$ flux.
As a result, the frequency obtained is $<$ 10$^{-5}$ for season 1 and season 4, $2 \times 10^{-1}$ for season 2.
For season 1 and 4, even when we injected a signal with the same variation as season 3, the 130-day period was not detected as a significant peak. This implies that the absence of the 130-day period in those seasons may be attributed to fewer data. 
However, for season 2, when we injected a signal with the same variation as season 3, the 130-day period could be detected. The 130-day period might not exist in season 2, regardless of the number of data.

It is conceivable that a variation with the same 130-day period exists constantly, but may not always be detectable due to its temporal absence. It may also indicate the on/off nature of the cycle. The temporal absence or unusual behaviors in activity cycles have been suggested for a few Sun-like stars, including the Sun. The Sun's 11-yr cycle, observed by the MWO HK project over 50 years (\cite{baliunas:1995}), exhibited peculiarities during solar cycle 22-23 (1985-2001). While the solar cycle minimum around 1985 was clearly seen, the subsequent minimum in 1996 was missing in certain activity indicators, such as the Na I D line (\cite{livingston:2007}). Also, a paucity of sunspots and a delay in the expected start of solar cycle 24 were reported (\cite{kilcik:2009}). 
In addition to the Sun, \citet{egeland:2015} reported that 58 Eri (HD30495, G1V) exhibits a quasi-periodic nature of short-period variations (1.7-yr). 
As mentioned earlier, $\epsilon$ Eri might also show temporal absence of the shorter cycle. 

The quasi-periodic cycles observed in the Sun-like stars can be influenced by various factors. For the Sun and $\epsilon$ Eri, unusual behaviors of the cycles may be associated with shorter or longer activity variations, such as Sun's 80-yr Gleissberg cycle and $\epsilon$ Eri's 12.7-yr cycle (\cite{richards:2009,metcalfe:2013}). However, \citet{egeland:2015} suggested that there may be no relationship between short and long periodicities in 58 Eri. 
Short and long cycles could interact in complex ways or be fundamentally different in nature. 
As any longer activity variations have not been reported in $\alpha$ CMi thus far, the nature of the short cycle and its temporal absence or existence remains unclear.
Nevertheless, we suggest that this specific behavior in the H$\alpha$ line represents a feature of the Sun-like magnetic activity cycle.
To better understand the nature of the shorter cycle and its possible relationship with longer cycles, further monitoring will be necessary.

\subsubsection{$\beta$ Vir}
The GLS periodogram of $\beta$ Vir in Figure \ref{figA1_1} (A) does not show any significant peaks with FAP $<$ 0.1\%. However, in the detrended data with the linear trend, a significant peak in the periodogram power was found at 530 days with FAP = $2 \times {10^{-6}}$ (Fig. \ref{figA1_1} (A)).
It is likely that the 530-day period is not influenced by the long-term trend. We estimated the error range of the detected period to be 530.5 $\pm$ 8.6 days.

\subsubsection{$\pi$$^{1}$ UMa}
In Figure \ref{fig3} (e), the integrated residual H$\alpha$ flux of $\pi$$^{1}$ UMa exhibits a distinct variation. Since no long-term trends were identified by the F-test, we performed the GLS periodogram analysis. The result is shown in Figure \ref{figA1_1} (B). The most significant peak was found at 560 days (f = 0.0018 day$^{-1}$) with an approximate FAP = 0.1\%. 
While it is possible that the observing time span and data points may not be sufficient to determine a clear magnetic activity cycle in this observation, we suggest that the 560 days is unlikely to be induced by a long-term trend and may represent an activity cycle with a significant FAP.

\subsubsection{$\beta$ Com, $\kappa$$^1$ Cet}
For $\beta$ Com and $\kappa$$^1$ Cet, we did not detect the estimated shorter cycles of 2.8 years (1022 days) and 365 days from the previous research (\cite{brandenburg:2017}), respectively (Fig. \ref{figA1_2} (C),(D)). For $\beta$ Com, a peak at 12.3 days (f = 0.0813 day$^{-1}$), close to the rotational period, became significant after removing the long-term trend.

We estimated the detection limit for the shorter cycle in the two stars by the simulation analysis. 
For $\beta$ Com, the amplitude of the H$\alpha$ variability with a frequency of more than 0.99 obtained is 1.25\% for the possible 1022-day cycle. 
The 1.7\% variability is over the detection limit for the possible 1022-day period in this observation, however, the estimated shorter cycle of 1022 days for $\beta$ Com was not detected. 
For $\kappa$$^1$ Cet, the amplitude of the H$\alpha$ variability with a frequency of more than 0.99 is 5.7\% for the possible 365-day cycle. 
In addition, the amplitude of the H$\alpha$ variability with a frequency of less than 0.01, representing the amplitude at which the possible 365-day cycle of $\kappa$$^1$ Cet may not be detected, is 2.6\%.
The obtained 2.5\% variability in this observation is not within the range of detection for the possible shorter period, suggesting why the estimated shorter cycle of 1 year for $\kappa$$^1$ Cet has not been detected.

\vskip\baselineskip
In our periodogram analyses of short-term variations for each star, we found peaks in the magnetic activity variability related to rotational periods in $\beta$ Com and $\epsilon$ Eri, both exhibiting significant power. $\kappa$$^1$ Cet and $\pi$$^{1}$ UMa showed weaker power associated with rotational periods. For 61 Vir, a peak at 37 days (f = 0.0270 day$^{-1}$) was implied, closely resembling its planetary orbital period (61 Vir c, 38 days, \citet{vogt:2010}), albeit with weak power. 
Also, $\eta$ Cep exhibited the highest peak at 163 days (f = 0.0061 day$^{-1}$), close to the possible planetary orbital period reported by \citet{nelson:1998}, but it is shown with very weak power.
These weak power levels in the short period range might be explained by factors such as short lifetimes of faculaes or variations in their distribution. Additionally, there is the possibility that large RV variations can shift the H$\alpha$ flux, potentially leading to fake variability. We confirmed that no significant differences beyond the error bar range in the integrated residual H$\alpha$ flux were observed for integration intervals wider than the 1 \AA.

Recent research, such as studies by \citet{parke-loyd:2022,hamaguchi:2023}, has highlighted the importance of understanding the properties of the prominent activity, such as flares or possible influences on exoplanets for the nearby bright Sun-like stars like $\epsilon$ Eri or $\kappa$$^1$ Cet. Investigating these Sun-like stars is crucial as they are likely to exhibit solar-like magnetic activity. Our H$\alpha$ results provide significant insights into the cyclic activity behaviors of these Sun-like stars.

\subsection{Short-term magnetic activity cycle in F-type stars}
\begin{figure*}[h]
 \begin{center}
  \includegraphics[width=100mm]{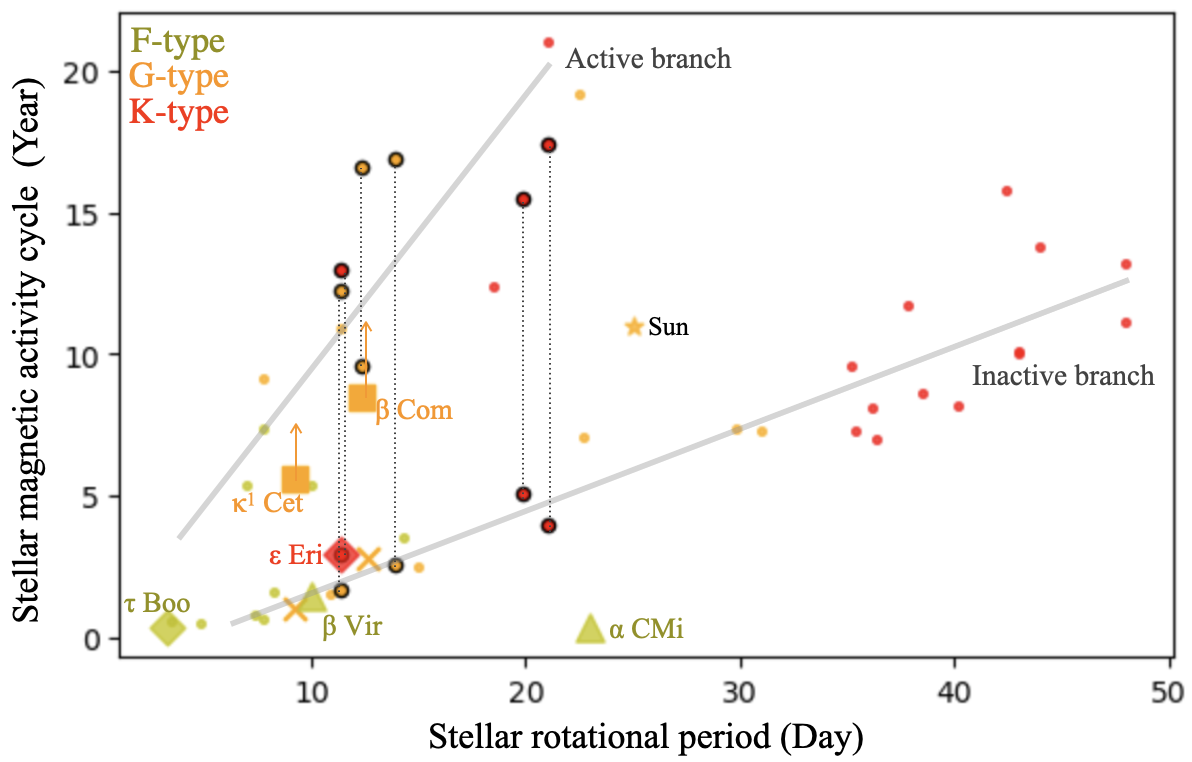}
 \end{center}
 \caption{The magnetic activity cycle versus the rotational period for F- (light green), G- (yellow), and K-type (red) stars. The data are taken from \citet{brandenburg:2017} and \citet{mittag:2019}. The two branches represent the active and inactive branch, respectively. The symbols with black edge, connected with a vertical dotted line, indicate the stars with two cycles. The diamond symbols indicate the target stars with the confirmed cycle in the H$\alpha$ observations ($\tau$ Boo, $\epsilon$ Eri). The triangle symbols indicate the target stars with the suggested cycle in the H$\alpha$ observations ($\beta$ Vir, $\alpha$ CMi). The square symbols with arrows indicate the targets implying possible longer cycles in the H$\alpha$ observations ($\beta$ Com, $\kappa$$^1$ Cet). The cross symbols indicate the expected shorter cycles ($\beta$ Com, $\kappa$$^1$ Cet) by \citet{brandenburg:2017}.}\label{fig10}
\end{figure*}

The relatively short time-scale magnetic activity cycle  appears to be a trend in F-type stars.
\citet{brandenburg:2017} proposed a spectral type dependence between the magnetic activity cycle and rotational period for the Sun-like stars, with distinct branches for active and inactive stars. 
The short-term activity cycles suggested in F-type stars fall within the inactive branch.
In this respect, we present the activity cycles of F-, G-, and K-type stars as a function of the rotational periods in Figure \ref{fig10}, combining updated data from previous studies (\cite{brandenburg:2017,mittag:2019}) with the results of HIDES-F H$\alpha$ observations. 
The samples of F-, G- and K-type stars, colored light green, yellow, and red, respectively, follow the general trend that the faster stars rotate, the shorter the magnetic activity cycle is. The magnetic activity cycles of F-type stars locate in a relatively short-term period range.

We previously reported an independent confirmation of the persistence of the short-term magnetic activity cycle ($\sim$ 123 days, light green diamond) in the H$\alpha$ variability for $\tau$ Boo (\cite{lee:2023}). Furthermore, we suggest the possibility of short-term activity cycles of 530 and 130 days for $\beta$ Vir and $\alpha$ CMi, respectively (light green triangles). These possible activity cycles in $\beta$ Vir and $\alpha$ CMi are within 1-2 years and shorter than those of G- or K-type stars with similar rotational periods. 
Our result, obtained through intensive monitoring with high cadence, supports the previous theoretical and observational lack of extremely short-term cycles in G- or K-type stars. The short-term activity cycle trend in F-type stars can be possibly related to the stellar intrinsic dynamo.

Meanwhile, the short period found in $\alpha$ CMi does not fall onto the inactive branch. It could be explained by its distinct characteristic, such as the star is located at edge of the main-sequence stars on the HR diagram.
Or, the period relying only on season 3 might not real for the entire cycle period. 
Another possibility is that short activity cycles result from variations in spot distributions or spot lifetimes ({\cite{basri:2020}}). To clarify this, it is important to investigate the absorption line profile caused by star spots in addition to continuous cycle observations with high cadence for $\alpha$ CMi.

\subsection{Co-existing activity cycles in G- and K-type stars} 
As described already, it is suggested that the magnetic activity cycle of the Sun-like stars tends to follow one of the two branches, and some G- and K-type stars with a shorter and a longer activity cycle of co-existing cycles are on both two branches (\cite{bohm:2007,brandenburg:2017}).
Also, there is the possibility that the branches show more scatters because cycle periods depend on stellar parameters like Rossby number. It is suggested that the active branch may actually consist of two sub-sequences (\cite{bohm:2007}), but it is not yet established.
We turn to Figure \ref{fig10} to discuss the H$\alpha$ results of our G- and K-type target stars.

$\beta$ Com and $\kappa$$^1$ Cet are expected to have a shorter cycle from the previous research.
From the results of $\beta$ Com in the H$\alpha$ observation, we found the possibility of a long-term variation beyond 8 yr (yellow square), and it is close to one of the detected cycles in $\beta$ Com. $\beta$ Com has been reported to show two cycles of 9.6 and 16.6-yr, however, the 9.6-yr cycle is not consistent with its expected shorter cycle (2.8-yr, yellow cross) and rather close to the active branch. As our H$\alpha$ observation for $\beta$ Com did not detect the expected shorter cycle even with likely sufficient amplitudes, it might support a sub-sequence in the active branch (\cite{bohm:2007}), unless the possible shorter cycle had a much smaller amplitude. It is not clear how stellar parameters affect the possibility of the sub-sequence to appear, so further observations of $\beta$ Com to determine co-existing cycles in the H$\alpha$ line could help establish different behaviors in the active branch.

The amplitude of a longer cycle would be greater than that of a shorter cycle (\cite{saar:2002,brandenburg:2017}), thus there is the possibility that a longer cycle is dominant among co-existing cycles. However, for $\epsilon$ Eri, the shorter cycle is clearly detected (red diamond), but its longer cycle was not implied in the H$\alpha$ line. Whereas for $\beta$ Com and $\kappa$$^1$ Cet, the shorter-term period was not found while the long-term trend is implied (yellow squares). Considering the longer cycles that were already found for these three stars, especially for $\beta$ Com, the longer cycle of $\epsilon$ Eri is not too long and therefore it can be implied during our observing time span. 
Thus, we speculate that a distinct shorter cycle could be dominantly detected. In our F-type stars, $\tau$ Boo and $\beta$ Vir, the clear variations of the short-term period appear dominantly as well (Fig. 5 of \citet{lee:2023} \& Fig. \ref{figA1_1} (A). 

The expected shorter cycle for $\beta$ Com and $\kappa$$^1$ Cet may have just too small amplitude to be detected. However, there is another possibility such as the temporal absence of the short-term variation. 
As described already, it is reported that the shorter cycle in $\epsilon$ Eri was not significant temporarily in the past. Also, one of our F-type stars, $\alpha$ CMi, has implied the possibility of the temporal absence/existence for the short-term variation. For $\alpha$ CMi, although the star showed the distinct short period in the specific season, the possible long-term trend was detected as the highest peak in the periodogram (Fig. \ref{fig7}). This could be because of the temporal absence/existence of the short period. Similarly, the long-term trend in $\beta$ Com and $\kappa$$^1$ Cet may be dominant just because their shorter cycle is off at present. The dominance or temporal absence/existence of the short cycle suggested by our observations could explain few cases of stars with detected co-existing cycles.

The common condition of stars with co-existing cycles in the two branches is an age range of 0.6 - 2.3 Gyr (\cite{brandenburg:2017}). Also, the activity level represented as log R'$_{HK}$ is around -4.75 (\cite{baliunas:1995}). For $\pi$$^1$ UMa, $\beta$ Aql, 61 Vir, $\eta$ Cep, and HD 102195, we could not find evidence of co-existing cycles. 
Given that either of their ages or activity levels are not satisfied with the conditions of co-existing cycles, the non-detection of any shorter periodic variations in our H$\alpha$ results for these stars may support the suggested conditions.
However, there is still the possibility to find evidence of co-existing cycles for some target stars, although we could not detect periodic variations in this observation. 
For $\pi$$^1$ UMa, considering its similarity of age and activity level to $\epsilon$ Eri, we can expect its shorter cycle of co-existing cycles. In practice, $\pi$$^1$ UMa shows a relatively distinct variation but with a slightly weak periodogram power, perhaps due to fewer data.
It suggests that conditions of co-existing cycles could be determined by considering more complicated factors related to stellar parameters.
Further monitoring would be essential to find robust periodic variations of $\pi$$^1$ UMa, to establish possible conditions of co-existing cycles as well.

\subsection{Impact of magnetic activity variability on exoplanet detection}
Based on the results in Table \ref{tbl3} and Section 5.2, we discuss the H$\alpha$ variability of the target stars focusing on exoplanet detection. Since the RV data with HIDES-F in almost the same seasons is not available at this stage, we investigated the previous RV results, the most recent ones possible, for the target stars. Owing to the possibility of changes of the magnetic activity variability with different epochs, we note that the H$\alpha$ results should only be considered as a possible suggestion. 

In the previous RV measurements for the four F-type stars (\cite{butler:1997,howard:2016,wittenmyer:2006,huber:2011}), the detected or suggested magnetic activity cycles in this observation are not implied. Also, for $\tau$ Boo and $\upsilon$ And, we did not find short-term periods related to their rotational period or planetary orbital periods in the H$\alpha$ observations (\cite{lee:2023}). 
It is suggested that the activity induced by the stellar rotation does not affect the H$\alpha$ variability for the F-type stars. Also, the impact of the long-term activity variability induced by the activity cycle might not be significant to the RV measurements.

For the five G-type stars, there is the possibility that the H$\alpha$ variability is affected by the stellar rotation and mimic a RV signal. 
Also, their long-term activity variations suggested by the H$\alpha$ line were not significant in the RV measurements (\cite{howard:2016}). 
Meanwhile, for 61 Vir, some activity variability may induce stellar noise to the RV for its Earth-mass planet.
In this observation, it is speculated that the G-type stars might be able to induce RV signals in various forms.

The possible impact of the H$\alpha$ variability on the RV measurements was different by individual stars for the three K-type stars.
For $\epsilon$ Eri, we found its rotational period in the periodogram, however, the 2.9-yr cycle was not found in the RV (\cite{hatzes:2000,benedict:2006}). 
For $\eta$ Cep, in addition to the weak periodogram power of the 163 days peak, as the existence of a hot Jupiter is not clear, we can not suggest the significance of the period. Nevertheless, there is the possibility that the magnetic activity variability might have induced the stellar noise to the RV.
For HD 102195, the RV measurements reported a 4.11-day orbit of its hot Jupiter (\cite{melo:2007}). Its substantial variability observed in the H$\alpha$ line could hinder searches for short-term periods.
The features of individual stars would be critical factors to make effects on the RV for the K-type stars. To understand the reasons for these differences, further investigation considering more various stellar parameters would be required.

Additionally, the results of this observation by the individual stars or spectral type are based on the observation with precision of few m/s.
With higher precision observations, the possible impacts can be detected.

\section{Summary}
We investigated the magnetic activity variability for 10 F-, G-, and K-type stars and features of the activity behaviors of the individual stars by intensive HIDES-F H$\alpha$ line monitoring during the last four years.
As a result, each star has shown its intrinsic magnetic activity variability. From the correlation between the H$\alpha$ variability and the Ca II H\&K line core emission, we suggest that the H$\alpha$ line can be used as an efficient indicator of the stellar magnetic activity. 
Also, a cyclic variation of 4 stars, 530-day of $\beta$ Vir, 130-day of $\alpha$ CMi, 560-day of $\pi$$^1$ UMa, and 2.9-year of $\epsilon$ Eri, was implied by the periodogram analyses. 

We also suggest that the activity variability represents different behaviors of the activity cycle depending on stellar spectral type. 
In addition to $\tau$ Boo from our previous report, we suggest a short activity cycle within 1-2 years for two F-type stars, $\beta$ Vir and $\alpha$ CMi. For the G- and K-type stars, the activity variation is implied to be longer than a few years. The clear cycle of 2.9-yr for $\epsilon$ Eri has been detected, and it is precisely consistent with its shorter cycle of co-existing cycles reported by the previous Ca II H\&K results. 
The expected shorter cycle periods for $\beta$ Com and $\kappa$$^1$ Cet were not detected in this observation, nevertheless, we set an upper limit of non-detection for the short-term period component of co-existing cycles. 

For the relationships between the stellar activity cycles and the rotational periods, we support the existence of extremely short periods of F-type stars in the inactive branch and the possibility of a sub-sequence in the active branch with intensive and high cadence monitoring. Also, our results for some G- and K-type target stars are consistent with the range of stellar age or activity level in which co-existing cycles would exist. However, since the existence of co-existing cycles could depend more complicated factors, such as combined influences from various stellar parameters, further observations are required to establish precise conditions.
Moreover, we found that $\alpha$ CMi has implied temporal absence/existence of the short-term variation, which may also affect detection of co-existing cycles.
It also emphasizes the significance of our intensive monitoring with high cadence.

With respect to the stars with a hot Jupiter, we could not find SPMI signals in HD 102195. In addition to our previous results of $\tau$ Boo and $\upsilon$ And, it is suggested that the magnetic activity variability in the H$\alpha$ observation is related to the stellar intrinsic activity rather than the existence of the hot Jupiter. 

We suggest that the magnetic activity variability can affect the RV differently by spectral type. We found the range of the amplitudes of the H$\alpha$ variability that may or may not induce a stellar noise to the RV with the precision of few m/s.

 \begin{ack}
This research is based on data collected with 1.88-m telescope at Okayama Branch Office, Subaru Telescope. The Okayama 1.88-m telescope is operated by a consortium led by Exoplanet Observation Research Center, Tokyo Institute of Technology (Tokyo Tech), under the framework of tripartite cooperation among Asakuchi-city, NAOJ, and Tokyo Tech from 2018. B.S. was partly supported by Grant-in-Aid for Scientific Research on Innovative Areas 18H05442 from the Japan Society for the Promotion of Science (JSPS), and by Satellite Research in 2017-2020 from Astrobiology Center, NINS. Y.N, was supportded by the JSPS KAKENHI Grant Number 21J00106.
Y.N. also acknowledge support from NASA ADAP award program Number 80NSSC21K0632. 

\end{ack}

\appendix
\renewcommand\thesection{\Alph{section}}
\section{Figures of periodogram analyses for target stars}
 \setcounter{figure}{0} 
 \renewcommand\thefigure{A.\arabic{figure}}
We present the GLS periodogram analyses for part of the target stars in Figure A.1. The upper panels show the periodogram results of all integrated residual H$\alpha$ flux time series data for each star. The middle panels shows the periodogram results after the long-term trends or periodicities were removed. The window function of the H$\alpha$ flux time series of each star was presented in the bottom panels. The two red lines represent the stellar rotational period and planetary orbital period, respectively. The blue line represents the short activity cycle from the previous reports. The green line represents the short activity cycle found in the integrated residual H$\alpha$ flux. 

\begin{figure*}[]
 \begin{center}
  \includegraphics[width=160mm]{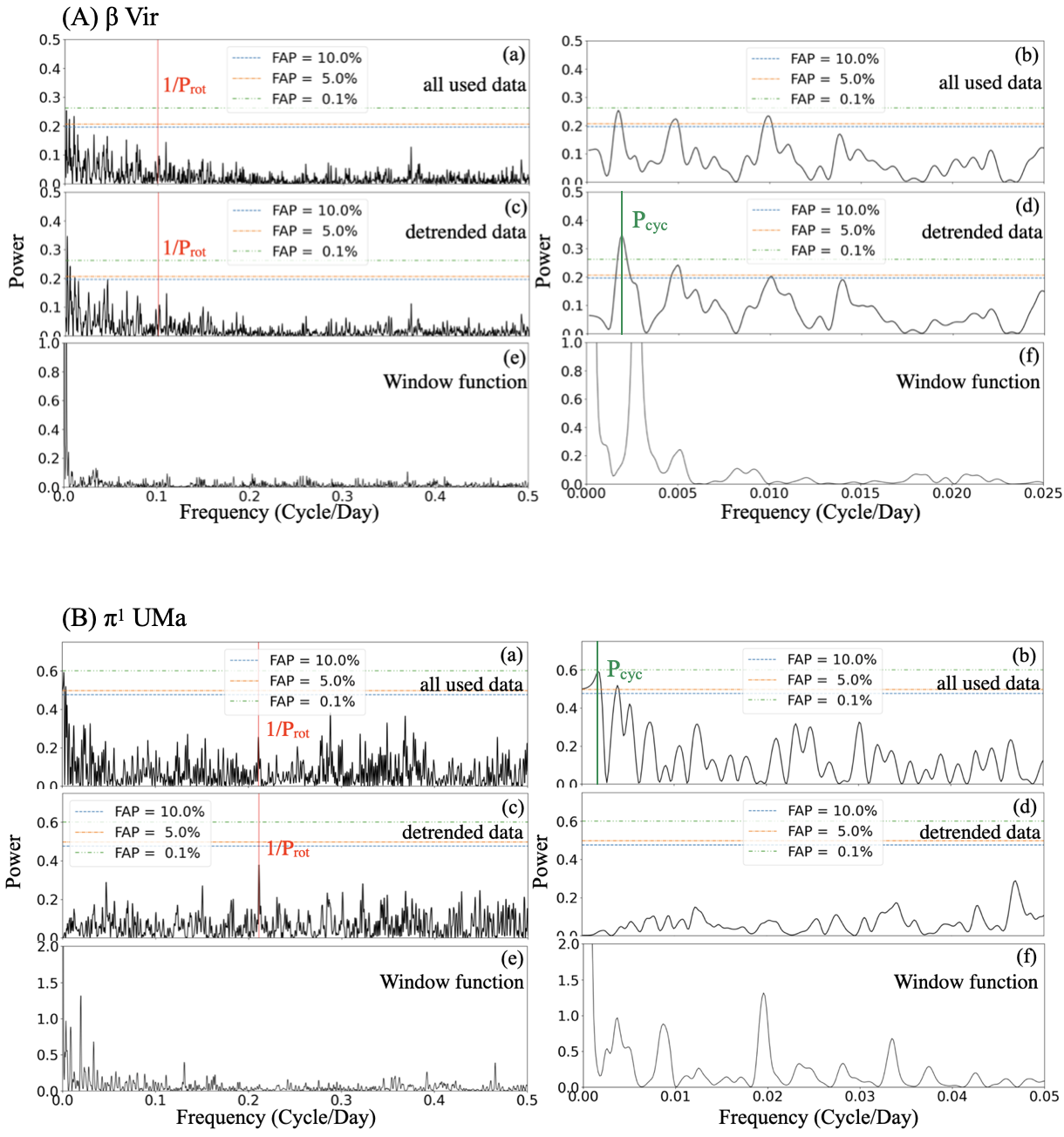}
 \end{center}
 \caption{(a) Periodogram of the integrated residual H$\alpha$ flux time series of each target star. (b) Enlarged portion of (a). (c) Periodogram of the detrended H$\alpha$ flux data of each target star. (d) Enlarged portion of (c). (e) Window function of the H$\alpha$ flux time series of each target star. (f) Enlarged portion of (e). The two red lines represent the stellar rotational period and planetary orbital period, respectively. The blue line represents the short activity cycle from the previous reports. The green line represents the short activity cycle found in the integrated residual H$\alpha$ flux.}\label{figA1_1}
 \end{figure*}

\addtocounter{figure}{-1}
 \begin{figure*}[]
 \begin{center}
  \includegraphics[width=160mm]{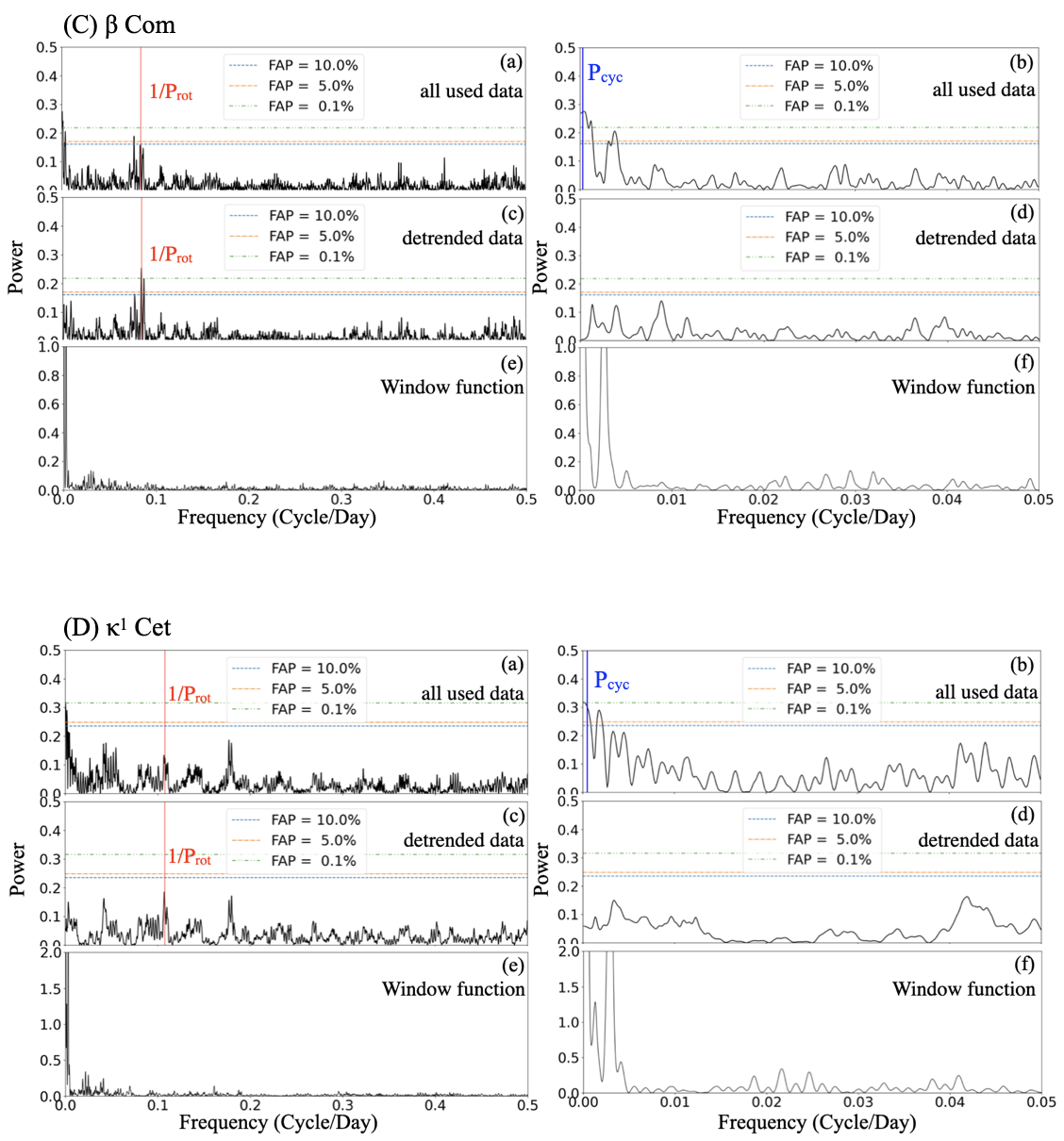}
 \end{center}
\caption{\label{}(Continued)}\label{figA1_2}
 \end{figure*}

\addtocounter{figure}{-1}
 \begin{figure*}[]
 \begin{center}
  \includegraphics[width=160mm]{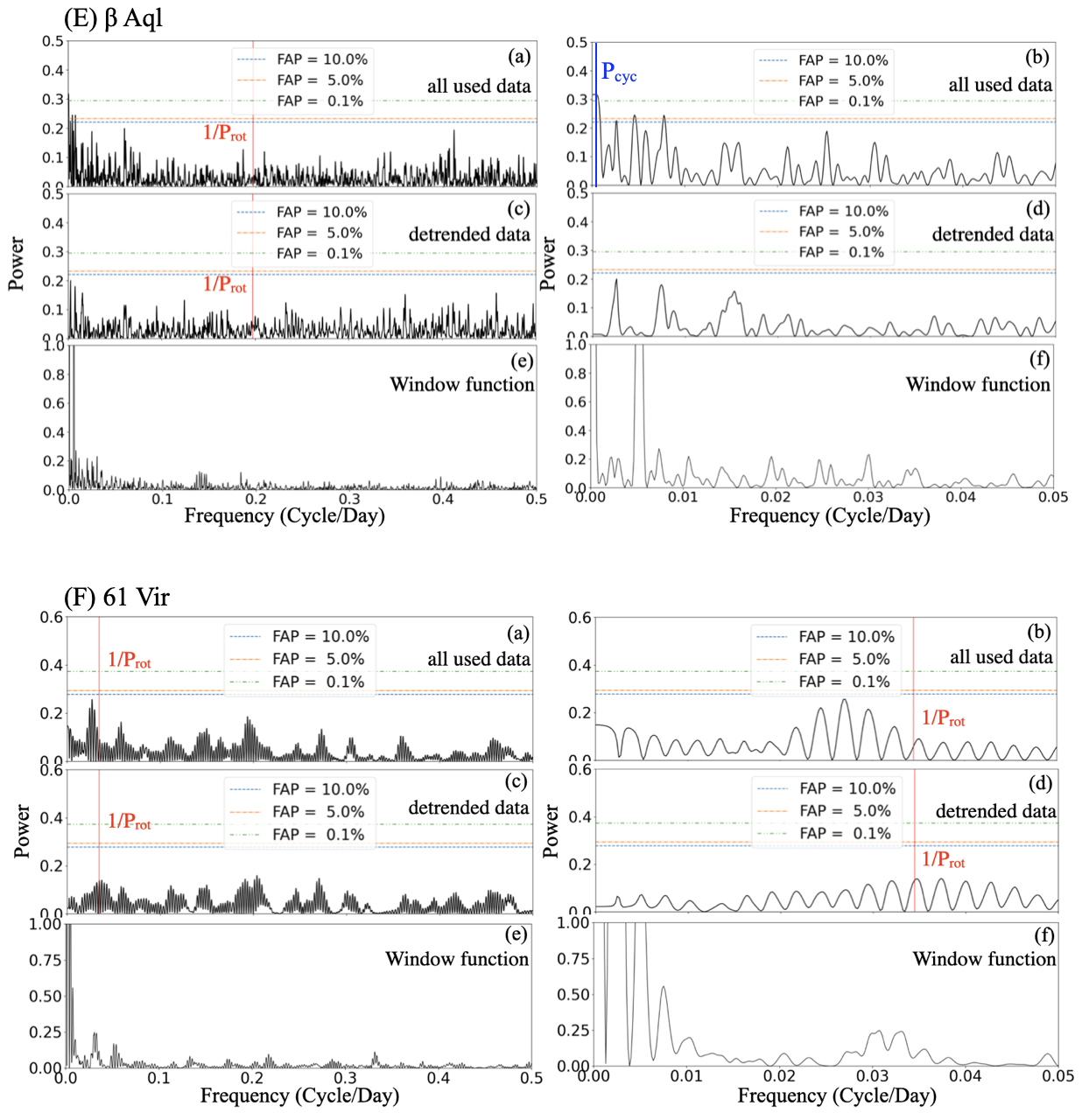}
 \end{center}
\caption{\label{}(Continued)}\label{figA1_3}
 \end{figure*}

\addtocounter{figure}{-1}
 \begin{figure*}[]
 \begin{center}
  \includegraphics[width=160mm]{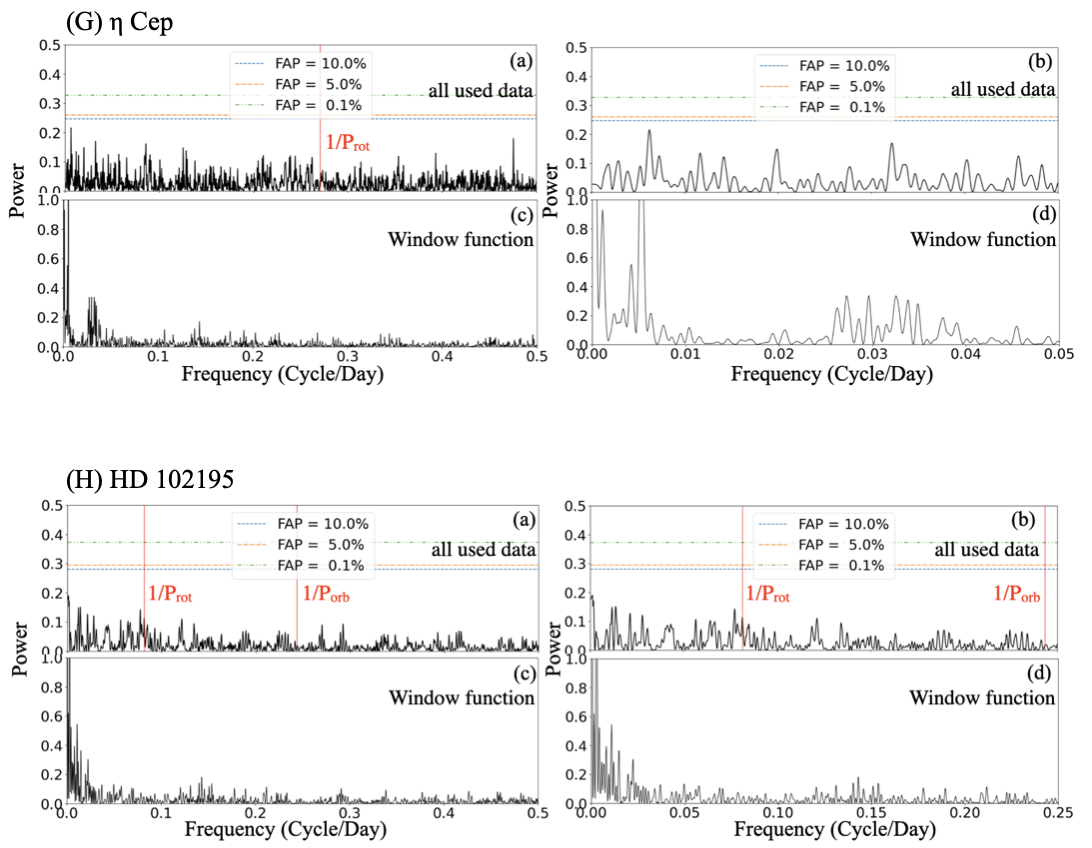}
 \end{center}
\caption{\label{}(Continued)}\label{figA1_4}
 \end{figure*}

\section{Figures of stellar parameter dependence of activity levels for target stars}
 \setcounter{figure}{0} 
 \renewcommand\thefigure{B.\arabic{figure}}
We present the correlation between several parameters and log R'$_{HK}$ for the target stars in Figure B.1.

\begin{figure*}[t]
 \begin{center}
  \includegraphics[width=140mm]{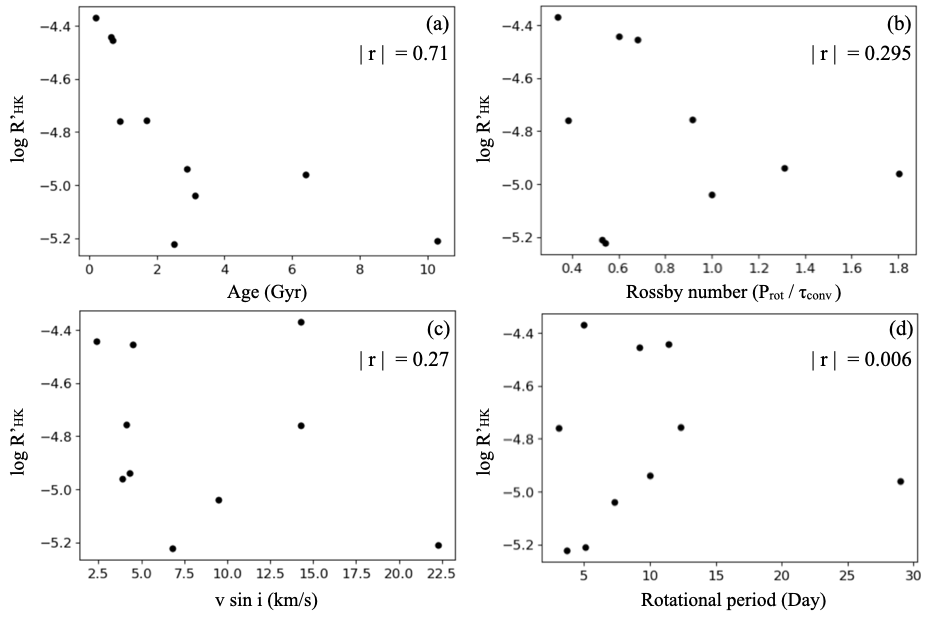}
 \end{center}
 \caption{Correlation between (a) stellar age, (b) Rossby number, (c) rotational velocity (\textit{v} sin \textit{i}), (d) rotational period, and log R'$_{HK}$. The correlation coefficient (r) is presented in each figure.}\label{figB1}
\end{figure*}

\end{document}